\documentclass[modern]{aastex63}

\usepackage{xcolor}

\renewcommand{\deg}{^{\circ}}
\usepackage{booktabs}

\submitjournal{ApJ}

\shorttitle{SFRs \& their properties via PSP dataset}
\shortauthors{Chen and Hu} 

\graphicspath{{FIGS/}}
\begin{document}

\title{Small-scale magnetic flux ropes and their properties based on in-situ measurements from Parker Solar Probe}
\correspondingauthor{Qiang Hu}
\email{qh0001@uah.edu}


\author[0000-0002-0065-7622]{Yu Chen}
\affiliation{Center for Space Plasma and Aeronomic Research (CSPAR), The University of Alabama in Huntsville, Huntsville, AL 35805, USA}

\author[0000-0002-7570-2301]{Qiang Hu}
\affiliation{Center for Space Plasma and Aeronomic Research (CSPAR), The University of Alabama in Huntsville, Huntsville, AL 35805, USA}
\affiliation{Department of Space Science, The University of Alabama in Huntsville, Huntsville, AL 35805, USA}


%
%
%
%
%
%
%



\begin{abstract}
We report small-scale magnetic flux ropes via the Parker Solar Probe in situ measurements during the first six encounters and present additional analyses to supplement our prior work in \citet{chen2021}. These flux ropes are detected by the Grad-Shafranov-based algorithm with the duration and scale size ranging from 10 seconds to $\lesssim$1 hour and from a few hundred kilometers to 10$^{-3}$ au, respectively. They include both static structures and those with significant field-aligned plasma flows. Most structures tend to possess large cross helicity, while the residual energy distributes in wide ranges. We find that these dynamic flux ropes mostly propagate anti-sunward, with no preferential sign of magnetic helicity. The magnetic flux function follows a power law and is proportional to scale size. We also present case studies showing reconstructed two-dimensional (2D) configurations, which confirm that the static and dynamic flux ropes have the common configuration of spiral magnetic field lines (also streamlines). Moreover, the existence of such events hints at the interchange reconnection as a possible mechanism to generate flux rope-like structures near the Sun. Lastly, we summarize the major findings and discuss the possible correlation between these flux rope-like structures and turbulence due to the process of local Alfv\'enic alignment.
\end{abstract} 


\keywords{Solar wind --- Astronomy data analysis --- Interplanetary turbulence --- Solar magnetic reconnection --- Solar magnetic fields}


%

\section{Introduction}
From the solar corona to the interplanetary space, the magnetic field lines are ubiquitous, which can stretch, twist, and reconnect. As a result of such processes, many structures are present in the solar wind. Magnetic flux rope is one of them, whose configuration consists of helical field lines. The traditional concept of the magnetic flux rope refers to a quasi-static structure, i.e., with almost no remaining plasma flow as viewed in a frame moving with the structure. The identification of magnetic flux ropes has been carried out for a wide range of scales, e.g., in duration from several minutes to days, at different heliocentric distances \citep{Cartwright2010,Chen2020}. In the study of \citet{Chen2020}, we reported thousands of small-scale magnetic flux ropes (SFRs) with the monthly occurrence rate of over two hundreds at both 1 au and $\geqslant$ 3.5 au via the ACE and Ulysses spacecraft measurements (see also, \citet{Chen2019}). The most distant observation of SFRs is obtained via the two Voyager spacecraft in situ measurements. The spacecraft traversed a dozen small structures that have duration of fewer than 9 hours at 9.57 au. Questions arise after finding those flux ropes at such distant places: what kind of variations or evolution will happen to them? Are those flux ropes originating from the Sun?

The large-scale counterparts of SFRs, usually classified as magnetic clouds, are from a subset of interplanetary coronal mass ejections (ICMEs). With the help of the coronagraph instrument, one can definitely observe the fact that near the Sun, CME usually has a clear expansion after being ejected from the Sun and propagates into the interplanetary space intercepting one or more spacecraft. However, many basic questions concerning SFRs remain. For instance, it is uncertain whether the SFR expands similarly to CMEs. On one hand, the flux rope merging process does cause SFR to increase in scale size. The observational result has confirmed this finding that two flux ropes merge into a larger structure \citep{Zheng2017}. Consequently, such a merging process due to magnetic reconnection leads to an increase in the toroidal flux, while the poloidal flux remains unchanged \citep{Fermo2011}. However, one should notice that what we rely on is mostly one-point observation, which delivers SFR information via the time-series data. Uncertainty exists in any analysis due to such a limitation on untangling spatial-temporal ambiguities. Therefore, we adopt an approach combining statistical analyses with individual case studies safeguarded by a set of quantitative metrics. It is also imperative to expand the analysis to additional spacecraft datasets in order to further examine the properties of SFRs and to address the questions on their origin and evolution.

From our prior studies, the statistical properties of identified SFRs via several datasets from 0.3 to 8 au hinted that they may have multiple origins. First, the Sun, as the source of the whole solar system, is generally believed to be responsible for generating these structures, likely near the Sun's surface. The observational analyses provided certain evidence for this view. For example, the occurrence of SFRs at 1 au generally follows the variation of the sunspot number with a short delay \citep{Hu2018}. Furthermore, the macroscopic properties at different distances and latitudes are usually in accordance with the solar wind characteristics \citep{Chen2019}. On the other hand, the widely identified current sheet structure exists at flux rope boundaries and together the coexistence complies with the scenario of turbulence-generated structures at magnetohydrodynamics (MHD) scales \citep{Servidio2009,Greco2008,Pecora2021}. In addition, the non-Gaussian distribution of the probability density function of the axial current density from these observational analyses is also consistent with that of turbulence-generated quasi-2D structures \citep{Zheng2018}. Thus, turbulent reconnection can act as a possible mechanism to produce SFRs as well, especially at the local site. In retrospect, the theoretical mechanism owing to magnetic reconnection was proposed when the notion of SFR was first put forward \citep{Moldwin2000}. This process is possibly associated with instabilities, which are able to produce the magnetic island configuration in the simulations \citep{Drake2006b,Nykyri2004}, although it can be regarded as a fundamental mechanism applicable to different plasma regimes.

With the launch of the Parker Solar Probe (PSP) mission, our focus turns more to the inner heliosphere given its close approaching distance to the Sun. One of the major discoveries is the ``omnipresent'' existence of magnetic switchbacks \citep{Bale2019,Kasper2019,Horbury2020,Dudok2020}. The possible generation mechanisms of these spikes in both in-situ magnetic and plasma measurements include the interchange reconnection between open and closed magnetic field lines \citep{Yamauchi2004,Sterling2020,Zank2020b,Liang2021}. In particular, such reconnection was proposed as a mechanism to produce magnetic flux ropes in low corona \citep{Drake2020}. They suggested that a flux rope structure with field-aligned plasma flows can be generated in a unidirectional background field and survive over long distances. Such type of flux rope, if crossed by a spacecraft, can also result in magnetic field reversals as indicative of switchbacks. Our recent study in \citet{chen2021} has affirmed this overlapping of identified flux rope and switchback intervals. They are two circumstances: (1) the spike fully encloses the flux rope or vice versa, and (2) two intervals have a partial overlap. Since both structures were identified from a single spacecraft measurements simultaneously or sometimes successively, it is very likely for them to form via the same mechanism(s) or represent the manifestations of the same structure. Moreover, similar power-law distributions of the waiting times also hint at this suggestion. 

We have implemented an automated detection algorithm based on the Grad-Shafranov (GS) reconstruction method \citep{Hu2017GSreview,Hu2018} in the previous studies \citep{Chen2020b,chen2021} in which relatively low sample rate data from PSP were employed. The duration range was 5.6 minutes $\sim$ 6 hours. In this study, we perform the extended GS-based detection algorithm for shorter duration, i.e., starting at a few seconds ($\sim$ 1000 km in cross-section size at a distance of a few tens of solar radii), to time periods around the first six perihelia. This additional analysis significantly extends the spatial scales examined at close distances from Sun that better represent the inertia-range turbulence, and alternatively the corresponding granular or supergranular structures on the solar surface. Current analysis also yields improved statistics in terms of significantly enlarged event sample size. Another major finding in our recent works is the prevailing dynamic flux rope structures identified by PSP dataset. By ``dynamic'', we mean those flux ropes with the magnetic field lines still taking the twisted shapes, but also containing significant remaining plasma flows that are aligned with the local magnetic field. Preliminary statistical analyses reveal that such structures show no significant deviation from static flux ropes in terms of their magnetic field configurations and other properties for a limited sample of events \citep{chen2021}. In this study, we thus adopt a broad definition of the flux rope and combine the previously separately defined structures of flux rope (``FR") and flux rope with field-aligned flow (``FRFF") as one unified entity, i.e., SFR or sometimes flux-rope like structures, because they are all governed by one generalized GS-type equation to be described in Section~\ref{sec:method}.

This paper is organized as follows. In Section \ref{sec:method}, we briefly recap the process of the GS-based detection algorithm and list the searching criteria. In Section \ref{sec:result}, we present the overview of these structures in six encounters (E1-E6) and show the statistical analyses of some basic parameters, including the Wal\'en test slope, normalized cross helicity and residual energy, duration, and scale size, for the identified SFR intervals. We also examine the correlation between selected parameters and the poloidal magnetic flux per unit length. In Section \ref{sec:case}, we present selected case studies and confirm the findings in \cite{Drake2020} that some magnetic switchback and flux rope-like structures can coincide. Finally, we summarize our major findings and discuss the similarities and differences of flux rope-like structures with turbulence and their relation to dynamic Alfv\'enic alignments in Section \ref{sec:sum}.

\section{Method Based on the GS-type Equation} \label{sec:method}
In this study, we use the extended approach of the automated flux rope detection algorithm based on the original Grad-Shafranov (GS) equation \citep{Sonnerup1996,Hau1999,2006JGRAS}, describing the force balance between the Lorentz force and the gradient of the thermal pressure $p$ in a 2D geometry ($\partial/\partial z=0$ but $B_z\ne0$), i.e., $\nabla^2 A = -\mu_0dP_t/dA = -\mu_0d(p+B_z^2/2\mu_0)/dA$. As introduced in \citet{Hu2001,Hu2002,Zheng2018,Hu2018}, an SFR interval can possess a double-folding pattern between the inbound and outbound paths along the spacecraft trajectory. Such a pattern is represented by two $P_t~versus~A$ branches, where $P_t$ is the transverse pressure, and $A$ is the magnetic flux function all obtained from time-series data. The original GS-based algorithm automatically scans all data arrays to look for good-quality patterns, i.e., candidates of SFRs. 
Notice that all calculations are processed in the co-moving frame, i.e., the de Hoffmann-Teller (HT) frame \citep{Khrabrov1998} with a constant frame velocity. In such a frame, the $z$-axis, i.e., the axis of a flux rope, is obtained via a trial-and-error process in the program loop. In the plane perpendicular to the $z$-axis, the $x$-axis is determined by the projection of the spacecraft path, and the $y$-axis completes the right-handed coordinate system \citep{Hu2002}.
The detailed flowchart illustrating the logic flow of the algorithm can be found at \url{http://fluxrope.info/flowchart.html}. A full description of the implementation is given in \citet{Hu2018}.

The extended GS-based algorithm takes modified forms of $P_t$ and $A$, denoted $P'_t$ and $A'$, respectively, which was implemented in \citet{chen2021}, taking into account non-vanishing remaining flow. The single-valued relationship of $P'_t~versus~A'$ enables it to be applicable to those structures with the remaining plasma flows that are aligned with and proportional to the local magnetic field in a proper frame of reference. A new type of GS equation is developed \citep{Teh2018}:
\begin{equation}
\label{eq:eq1}
\nabla^2 A'=-\mu_0\frac{d}{dA'}\left[(1-\alpha)^2\frac{B^2_z}{2\mu_0}+(1-\alpha)p+\alpha(1-\alpha)\frac{B^2}{2\mu_0}\right], 
\end{equation}
where $\alpha=\langle M_A\rangle^2\approx Const$, and $\langle M_A\rangle$ is the average Alfv\'en Mach number, the ratio between the remaining flow and the local Alfv\'en velocity. This formulation is consistent with an alternative and more general formulation presented by \citet{2006JGRAS}, with $P'_t$ corresponding to the terms enclosed in the square brackets of the right-hand side of equation~(\ref{eq:eq1}).

\begin{table}
\begin{center}
\caption{Detection criteria of SFRs for the GS-based algorithm.}
\begin{tabular}{ccccc}
\toprule
Duration (seconds) & Wal\'en Test Slope & $\langle B \rangle$ (nT) & $\langle M_A\rangle$ & $|R|$\\
\midrule
10 $\sim$ 344 & $\leqslant$ 1.0 & $\geqslant$ 25 & $<$ 0.9 & $\geqslant$ 0.8 \\
\bottomrule
\end{tabular}
\label{table:criteria}
\end{center}
\end{table}

We use this extended formulation to search for the new double-folding pattern of $P'_t~versus~A'$, where
\begin{equation}
\label{eq:eq2}
A'(x,0)=-\int_0^x (1-\alpha)B_y(x',0) dx',
\end{equation}
a line integral of the measured magnetic field component $B_y$ along the spacecraft path at $y=0$.
Similar to \citet{Hu2018}, Table \ref{table:criteria} lists the criteria for this algorithm. It is implemented via a set of sliding windows ranging from 10 to 344 seconds in size (range of duration of identified SFR intervals). We use the Wal\'en test slope to evaluate the Alfv\'enicity of a structure. It is calculated via the linear regression between the three components of $\textbf{V}_{rel}-\textbf{V}_{HT}$ and $\textbf{V}_A$, where $\textbf{V}_{rel}$ is the relative proton bulk velocity which takes spacecraft velocity into account (both given in an inertia frame), $\textbf{V}_{HT}$ is the velocity of the HT frame, and $\textbf{V}_A$ is the local Alfv\'en velocity. 
Since we do not distinguish the static flux rope from the Alfv\'enic ones in this study, the threshold of the Wal\'en test slope is relaxed to be 1.0. Moreover, in order to eliminate small fluctuations, we set a limit on the field magnitude, i.e., 25 nT. Considering the applicability of the new GS equation (\ref{eq:eq1}) to avoid the singularity for $\alpha=1$, we also set $\langle M_A\rangle$ to be less than 0.9. Last but not least, the absolute value of the correlation coefficient $R$ between $\textbf{V}_{rel}-\textbf{V}_{HT}$ and $\textbf{V}_A$, is required to be $\geqslant$ 0.8 to indicate that the remaining plasma flow is well aligned with the local magnetic field, such that $\alpha\approx Const$ can be satisfied.

In addition, two auxiliary parameters are employed to evaluate the Alfv\'enicity, i.e., the normalized cross-helicity density $\sigma_c$ and the normalized residual energy density $\sigma_r$ \citep{Roberts1987b,Matthaeus1982,Bavassano1998}. These two quantities are approximated in the time domain and are calculated by the following equations:
\begin{equation}
\label{eq:eq4}
\sigma_c=2\langle\textbf{v}\cdot\textbf{b}\rangle/(\langle v^2\rangle+\langle b^2\rangle)
\end{equation}
\begin{equation}
\label{eq:eq5}
\sigma_r=(\langle v^2\rangle-\langle b^2\rangle)/(\langle v^2\rangle+\langle b^2\rangle), 
\end{equation}
where \textbf{v} represents the remaining flow velocity in the HT frame, \textbf{b} is the magnetic field in the Alfv\'en unit, and $\langle \cdot\rangle$ means the average within event interval. In addition to the Walén test slope, these two quantities can also specify the degree of the Alfv\'enicity. Generally, the high Alfv\'enicity is pronounced when $\sigma_c$ and $\sigma_r$ approach $\pm$1 and 0 respectively \citep{Bruno2013}. Moreover, depending on the polarity of the background magnetic field, the cross-helicity with large magnitudes usually indicates outward/inward propagating Alfv\'en waves. Meanwhile, we also obtain the extreme value of the poloidal magnetic flux function, $A_m$, residing in the array $A(x,0) = A'/(1-\alpha)$ from equation (\ref{eq:eq2}). As aforementioned, $A$ represents the magnetic flux function. Therefore, the quantity $A_m$ here refers to the difference in $A$ between the boundary and the center of the structure, thus the absolute value $|A_m|$ yielding the amount of poloidal magnetic flux per unit length. The sign of $A_m$ indicates the sign of magnetic helicity or the chirality of the SFR.

The data of the magnetic field and proton bulk properties are recorded by the FIELDS Experiment \citep{Bale2016} and the Solar Wind Electrons Alphas and Protons (SWEAP; \citet{Kasper2016,Case2020}) instrument suite, respectively. 
The magnetic field and plasma bulk parameters (for proton only, no electron data available), including velocity, number density, and temperature, are public data with the tag ``Only Good Quality'' and available on the NASA CDAWeb.
This study mainly focuses on periods where the high cadence encounter mode is on. During these time periods, the data resolution for plasma is usually about 0.873 seconds, while the cadence of the magnetic field is always less than 0.437 seconds. In this study, we downsample all data to a cadence of 1 second.

\section{Macroscopic Properties}\label{sec:result}

\begin{table}[ht]
\begin{center}
\caption{Detection Results during the first six PSP encounters.}
\begin{tabular}{lcccccc}
\toprule
PSP Encounters & E1 & E2 & E3 & E4 & E5 & E6\\
\midrule
Time Periods (days)& 12 & 12 & 8 & 14 & 16 & $<$ 8 \\
SFR Duration (seconds) & 10-2,605 & 10-1,205 & 10-3,697 & 10-343 & 10-2,633 & 10-1,793 \\
Event Counts & 1,003 & 1,466 & 850 & 1,459 & 820 & 243\\
\bottomrule
\end{tabular}
\label{table:result}
\end{center}
\end{table}

As listed in Table \ref{table:result}, the extended GS-based algorithm is applied to time periods around the first six PSP perihelia when the high cadence data are available. The time periods for detection start from 2018 October 31 to November 12, 2019 March 30 to April 11, 2019 August 23 to 31, 2020 January 23 to February 8, May 29 to June 14, and September 18 to October 1, for encounters E1-E6, respectively. The new detection is implemented for a total detection period of over two months. As aforementioned, the duration limit in the detection is set to be 10 $\sim$ 344 seconds. In order to acquire event characteristics on a wider range of scales, we also combine the current detection results with the SFR candidates that have duration from 337 seconds to $\sim$ 6 hours in the prior study \citep{chen2021}. During 70 days of detection periods, we totally identify 5,841 events including both static and dynamic flux ropes at heliocentric distances between 0.13 and 0.35 au. The duration and scale size of these events range from 10 to 3,697 seconds and from 3.99$\times 10^{-6}$ to 5.96$\times 10^{-3}$ au, respectively.
On average, the daily occurrence rate is about 103 events per day for E1-E4, which is fairly persistent. For E5-E6, the occurrence rate drops, which is ascribed to the lack of solar wind velocity data near perihelia. 

\begin{figure}[ht]
\centering
\includegraphics[width=1.0\textwidth]{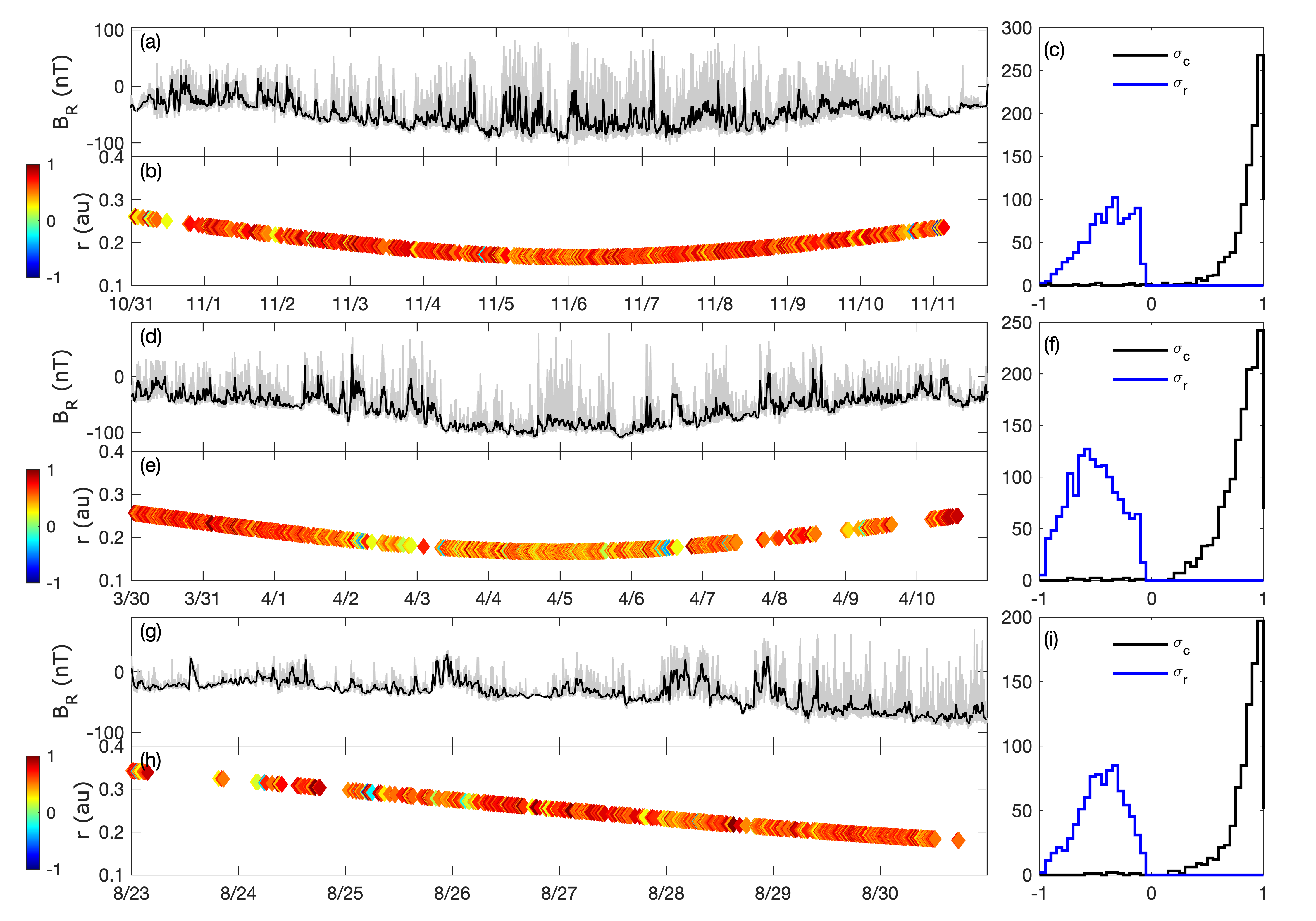}
\caption{Summary plot for the first three perihelia: (a-c) E1, 2018 October 31 to November 12, (d-f) E2, 2019 March 30 to April 11, and (g-i) E3, 2019 August 23 to 31. For each encounter, the panels are the time-series plot (first panel) of the radial magnetic field $B_R$, the corresponding heliocentric distance of each event (colored symbol) with the color representing the Wal\'en test slope as indicated by the colorbar (second panel), and the distribution of normalized cross-helicity $\sigma_c$ as well as normalized residual energy $\sigma_r$ (right panel). The radial magnetic field measurements with 1s cadence and 1250s running average are shown by gray and black curves, respectively.}\label{fig:E1-3}
\end{figure}

\begin{figure}[ht]
\centering
\includegraphics[width=1.0\textwidth]{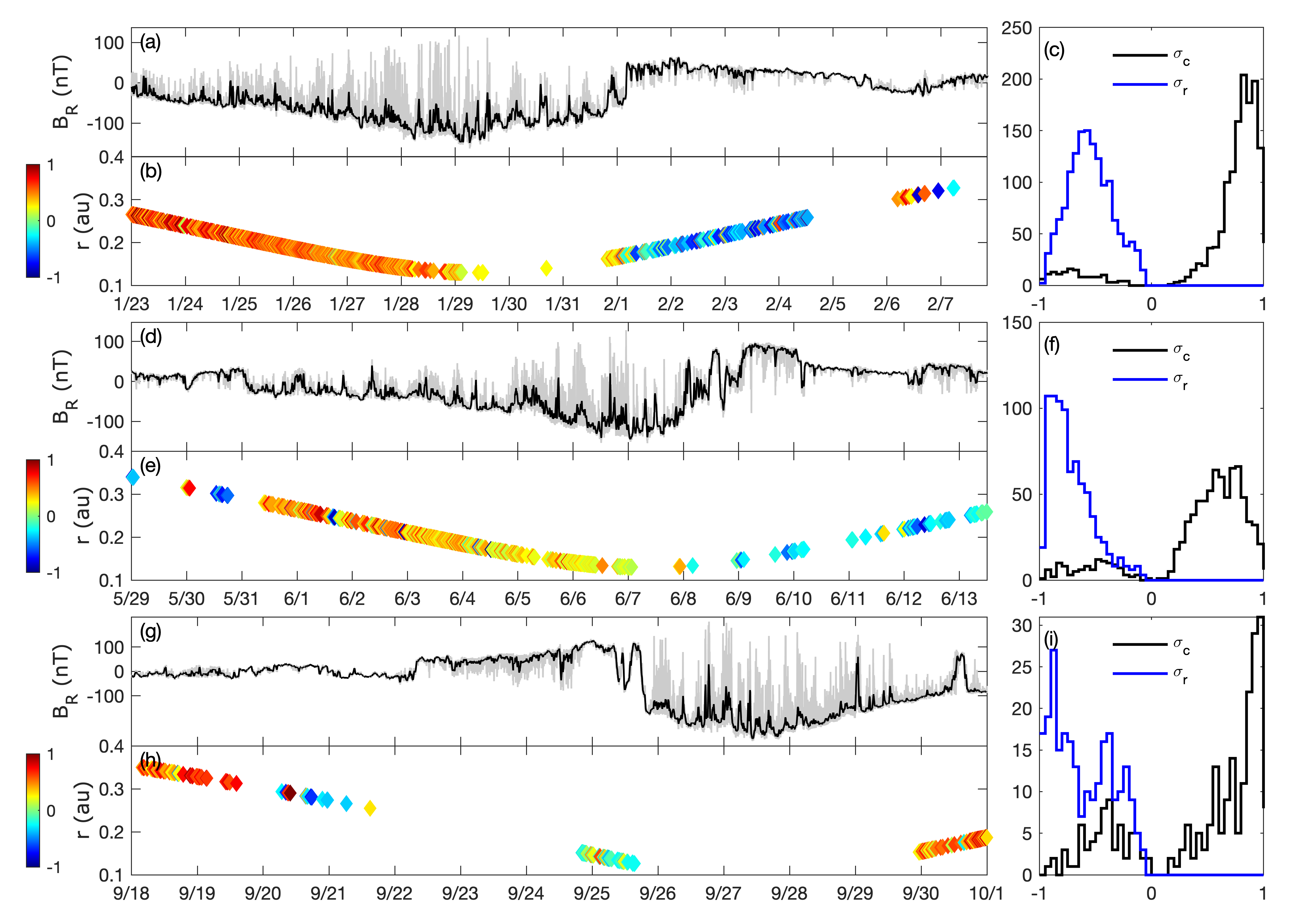}
\caption{Summary plot for E4-E6: (a-c) E4, 2020 January 23 to February 8, (d-f) E5, 2020 May 29 to June 14, and (g-i) E6, 2020 September 18 to October 1. The format follows that of Figure \ref{fig:E1-3}.}\label{fig:E4-6}
\end{figure}

Figures \ref{fig:E1-3} \& \ref{fig:E4-6} present time-series plots in encounters E1-E6 for the radial magnetic field $B_R$, radial distance $r$ of PSP, distributions of the Wal\'en test slope, and normalized cross helicity $\sigma_c$ as well as residual energy $\sigma_r$. In each figure, panels (a,d,g) show the radial magnetic field $B_R$ and its 1250-second running average. Obvious enhancements of the magnetic field intensity can be seen when the PSP approaches the perihelia with decreasing radial distance. Such a process is sometimes followed by a change of magnetic polarity accompanied with the corresponding change in the electron pitch angle distribution (ePAD; not shown), indicative of the heliospheric current sheet (HCS) crossing \citep{Whittlesey2020}. Complete crossings of HCS are pronounced in encounters E4 and E5 \citep{Phan2021}, and possibly E6 as well. In the first three encounters, one polarity, mostly negative, dominates during the time periods in Figure \ref{fig:E1-3}. 
Panels (b,e,h) present the values of the Wal\'en test slopes for identified SFRs as a function of time and the corresponding heliocentric distances. Although values of Wal\'en slopes range from -1 to 1 as indicated by each colorbar, 99\% of events in E1-E3 own positive slopes (995/1,003 in E1, 1,452/1,466 in E2, and 841/850 in E3). 
On the other hand, such a ratio changes in E4-E6 since the radial magnetic field turns to be positive in the outbound paths. According to \citet{Phan2021}, three HCS crossings in E4 and E5 start from 2020 February 1, 04:03:46 UT, 2020 June 8, 11:05:56 UT, and 15:40:45 UT, respectively. Before the HCS crossings, 98\% of events (1281/1288 in E4 and 705/737 in E5) have positive Wal\'en test slopes under the circumstance of the negative radial magnetic field. After a complete crossing when $B_R$ changes the sign, there are 84\% of events (134/171 in E4 and 79/83 in E5) possessing negative slopes with simultaneously positive $B_R$. It seems that the PSP also completed an HCS crossing in E6, although it was not covered in \citet{Phan2021}. In Figure \ref{fig:E4-6}(g-h), there are negative (positive) Wal\'en slopes mainly associated with the positive (negative) $B_R$, although event counts have decreased significantly due to data gaps.
In Figure \ref{fig:E1-3}(c,f,i), distributions of normalized cross helicity $\sigma_c$ and normalized residual energy $\sigma_r$ are displayed in black and blue lines respectively. Distributions of $\sigma_r$ show cluster within the negative value range from -1 to 0. This corresponds to one of our detection criteria, e.g., $\langle M_A\rangle < 0.9$, and demonstrates that the kinetic energy within these flux rope-like structures is modestly smaller than the magnetic energy. Results in encounters E2, E3, and E4 seem to have random $\sigma_r$ values, while results in E1 and E5 tend to have skewed values toward $\sim$ 0, and -1, respectively. Although distributions in E5 \& E6 may be limited by event counts, these values reflect that there seems to exist significant variability in Alfv\'enicity, mostly ranging from modest to high levels as judged by the values of $\sigma_r$.
Distributions of $\sigma_c$ in all encounters are asymmetric. The positive signs of $\sigma_c$ are dominant in the first three encounters, while minority events have negative signs. Again, it indicates that the positive values appear to take place in the background field of mostly negative polarity. When the background field polarity changes from being negative to positive, more negative values arise. Such a change is seen in E4-E6. Clearly, these changes are due to the HCS crossings during E4-E6, which coincide with the change of the sign of $\sigma_c$. The overall tendency indicates that the cross helicity $\sigma_c$ is largely positive in a background field of negative polarity ($B_R$ component), and vice versa. Such a correspondence implies that most structures, if they can, are propagating outward (away from the Sun) \citep{Tu1995,Bruno2013,Zhao2020aa}, consistent with the finding in \citet{Parashar2020}.

\begin{figure}[ht]
\centering
\includegraphics[width=0.45\textwidth]{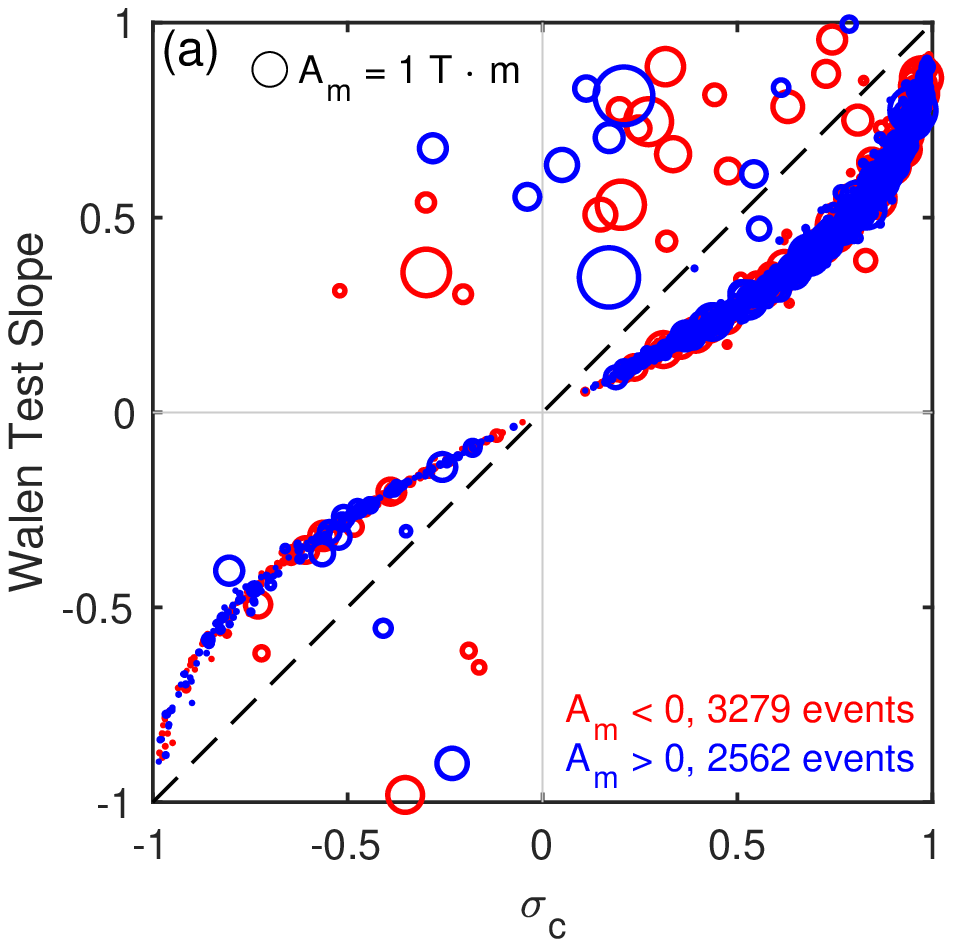}
\includegraphics[width=0.45\textwidth]{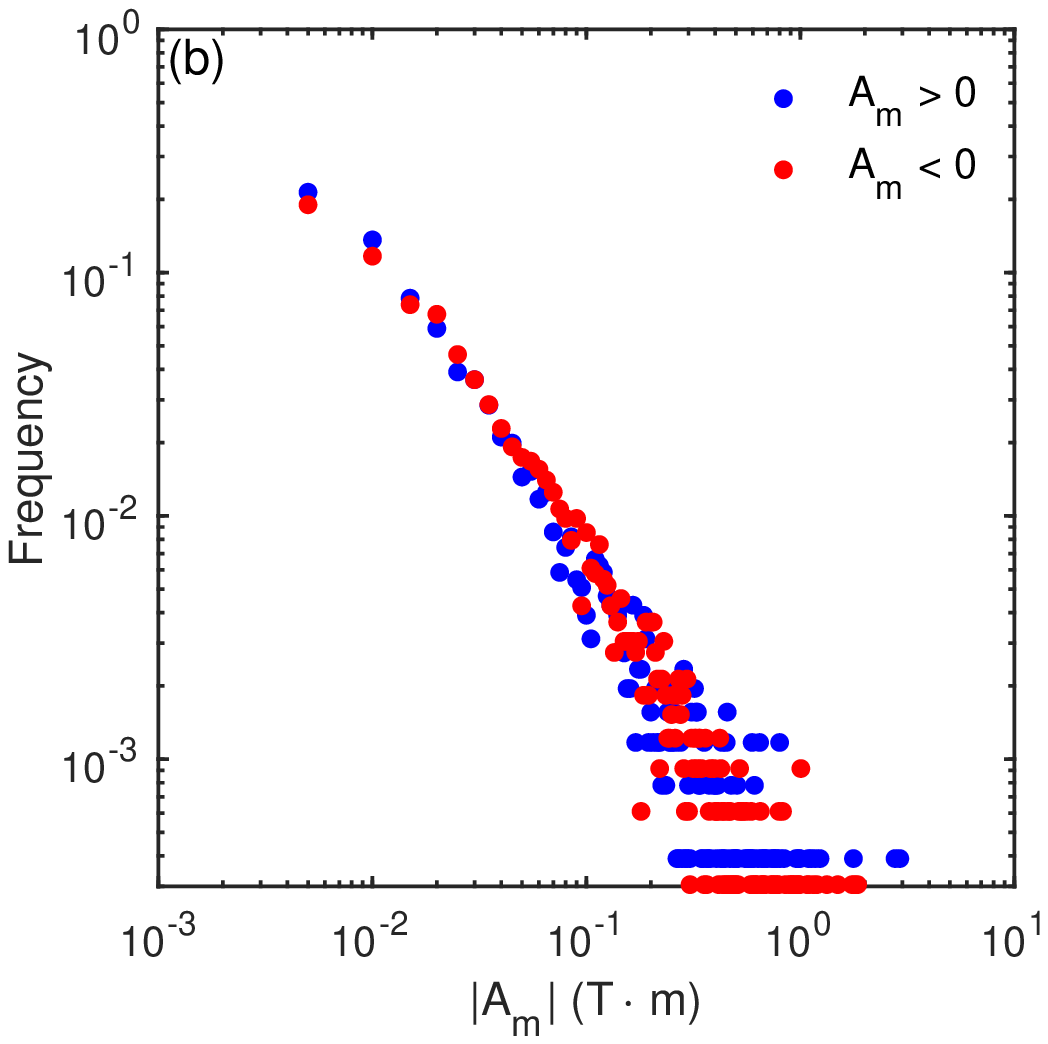}
\caption{Left panel: Distribution of the Wal\'en test slope versus the normalized cross helicity $\sigma_c$. Events that possess positive and negative signs of magnetic helicity are denoted by blue and red circles, respectively. The size of each circle is proportional to the magnitude of $A_m$. The circle in the top left is a reference symbol for $|A_m|$ = 1 T$\cdot$ m. The diagonal dashed line indicates where the two parameters are equal. Right panel: Distribution of $|A_m|$. Events that possess positive and negative signs of the magnetic helicity are denoted by blue and red dots, respectively.
}\label{fig:cross}
\end{figure}

We also compare the Wal\'en test slope with normalized cross helicity $\sigma_c$ as well as distribution of the poloidal magnetic flux of each SFR, i.e., $|A_m|$, in Figure \ref{fig:cross}. For each circle in Figure \ref{fig:cross}(a), blue and red colors denote the positive and negative signs of $A_m$, while the size of the circle represents its magnitude. The symbols are largely aligned with the diagonal line, indicating that the Wal\'en slope and $\sigma_c$ have the same sign and are also comparable in magnitude. 
It seems that these two quantities are connected intrinsically, which is expected since they all reflect the relation between the remaining flow and Alfv\'en velocities. However, the physical connection between these two quantities is unknown.
Additionally, combining with the distributions of $\sigma_c$, those events that possess positive (negative) Wal\'en slopes in the background of the negative (positive) $B_R$ also correspond to outward propagating SFRs (if they can). The ratio is 98.02\% in the E1-E5. Only 1.98\% of events have the same signs of the Wal\'en test slopes and $B_R$, which are possibly inward propagating structures. Events with positive/negative signs of $A_m$ are marked in blue and red respectively. Notice that the chirality or the sign of magnetic helicity is equivalent to the sign of $A_m$. Totally, there are 3,279 and 2,562 events possessing negative and positive magnetic helicity, respectively.
For a flux rope configuration, the positive or negative sign of helicity corresponds to right-handed or left-handed chirality, respectively. Although the numbers of events are different, there reveals no significant preferential distribution of the poloidal flux for events with positive and negative magnetic helicity, as seen in Figure \ref{fig:cross}(b). 
The overall distribution (for either $A_m>0$ or $A_m<0$) behaves like a power-law function.

\begin{figure}[ht]
\centering
\includegraphics[width=0.45\textwidth]{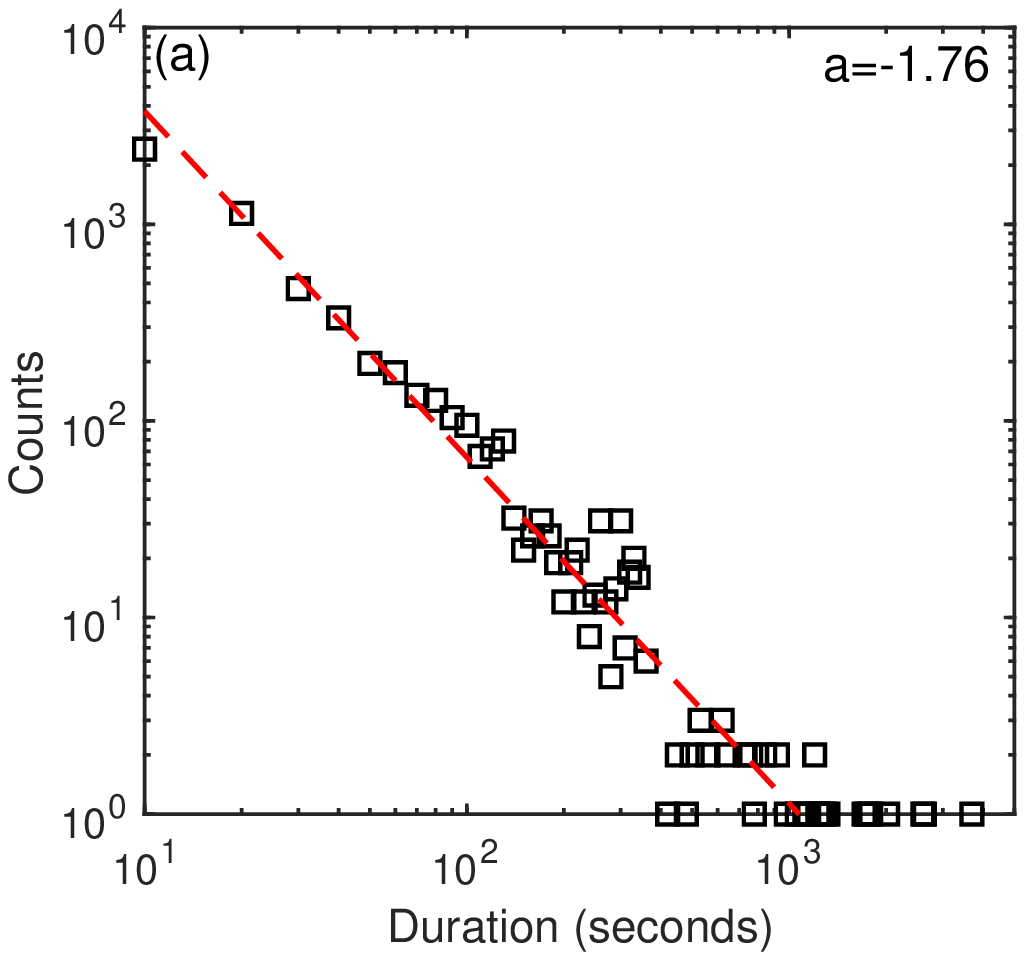}
\includegraphics[width=0.45\textwidth]{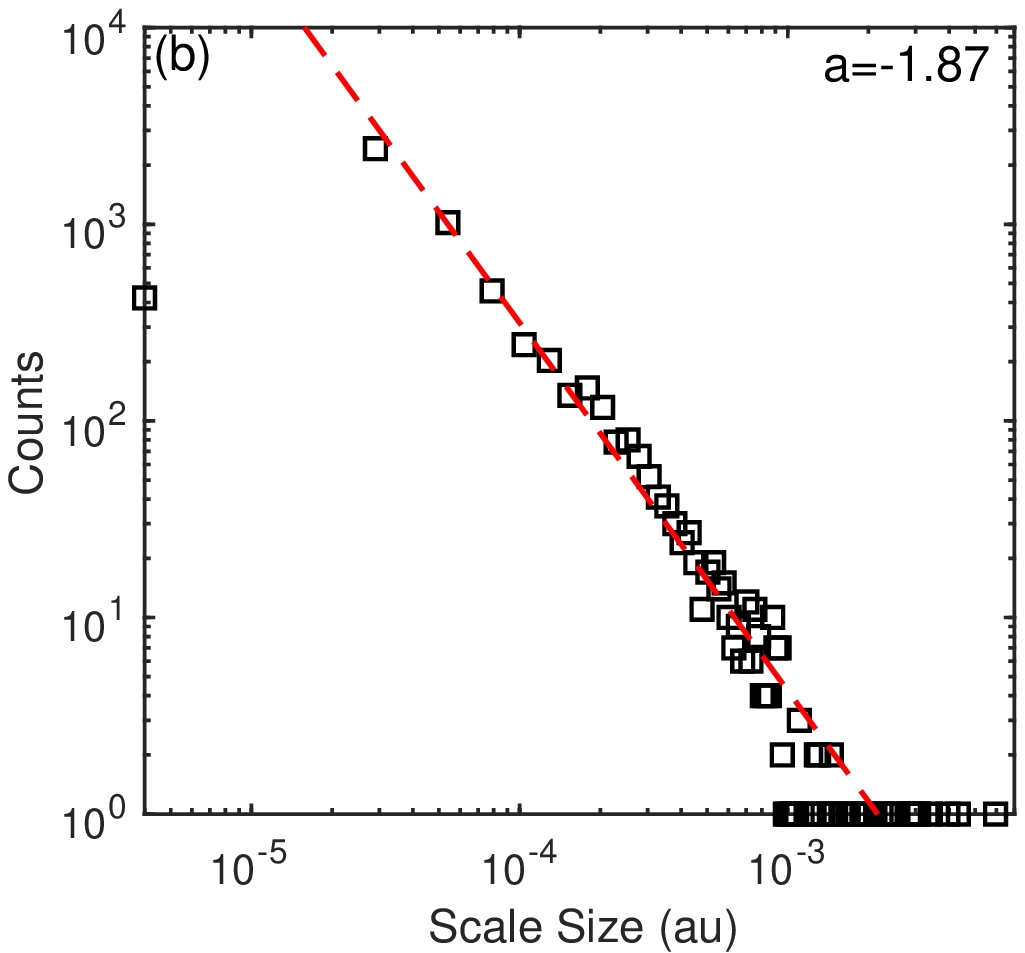}
\caption{Distributions of (a) duration and (b) scale size. The red dashed line is a power-law fitting function with the exponent $a$ indicated. 
}\label{fig:duration1}
\end{figure}

Figure \ref{fig:duration1} presents distributions of event duration and cross-sectional scale size. The duration measures the length of an event interval, while the scale size in this study is calculated by multiplying the $x$ component of $\textbf{V}_{HT}$ and the event duration. Although the event duration has been replenished to 6 hours, events with smaller duration and scale sizes still prevail. In the previous report \citep{Chen2020}, we found that distributions of these two parameters follow power-laws at different heliocentric distances, i.e., 0.3 $\sim$ 9 au. Such tendencies now extend to smaller scales and to smaller heliocentric distances. Each distribution approximately follows a single power-law function.
The power-law indices are around -1.8. We notice that \citet{Dudok2020} reported the power-law distribution of the duration of the magnetic switchback with indices falling within -1.4 and -1.6. Actually, their work includes lots of events that are shorter than ours, i.e., with duration down to 10$^{-2}$ s because they solely based their analysis on magnetic field data. Those events dominate and thus have significant effects on the power-law indices. 
In \citet{Dudok2020}, at the lower end of the distribution of duration, events under different thresholds of normalized deflection parameter follow a unified power law, then they start to deviate at tails. The scale of our events just corresponds to this deviating part. Therefore, a direct comparison is hard to achieve at the present time, although certain degree of similarity in terms of a power-law distribution in duration is seen.

\begin{figure}[ht]
\centering
\includegraphics[width=0.45\textwidth]{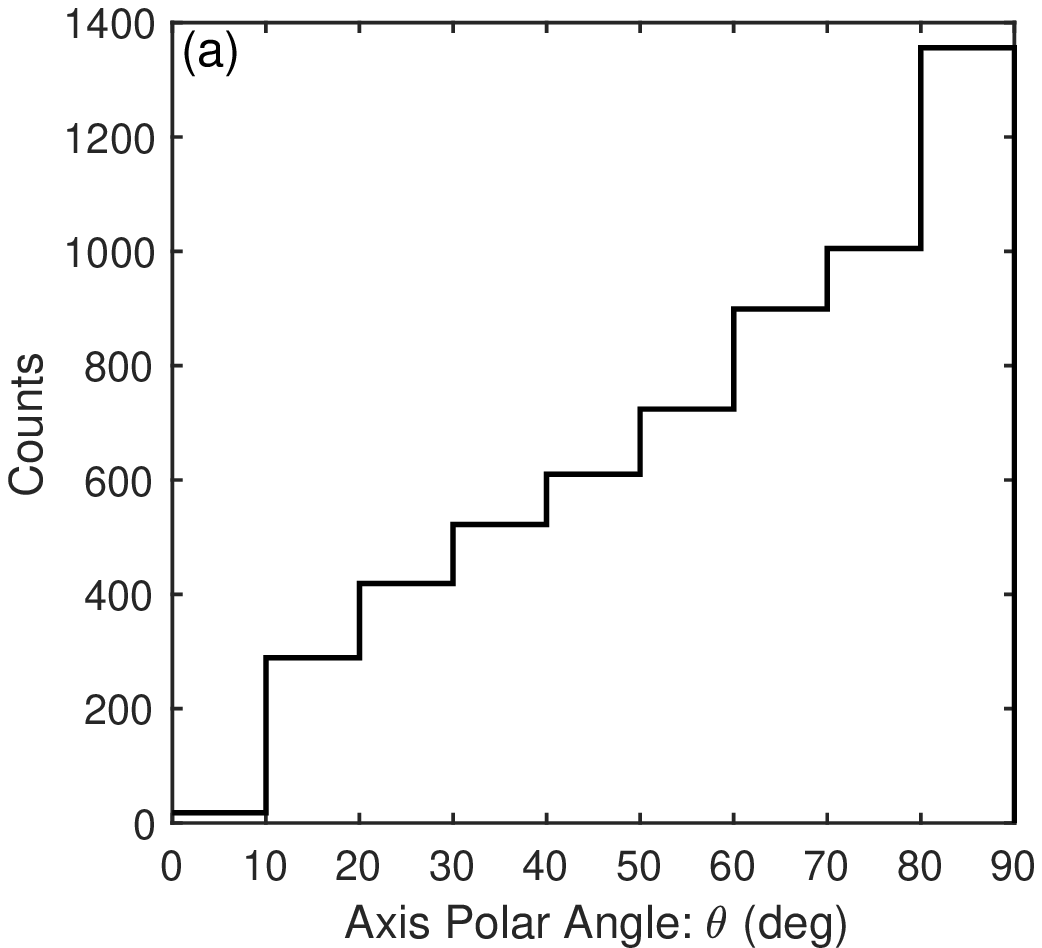}
\includegraphics[width=0.45\textwidth]{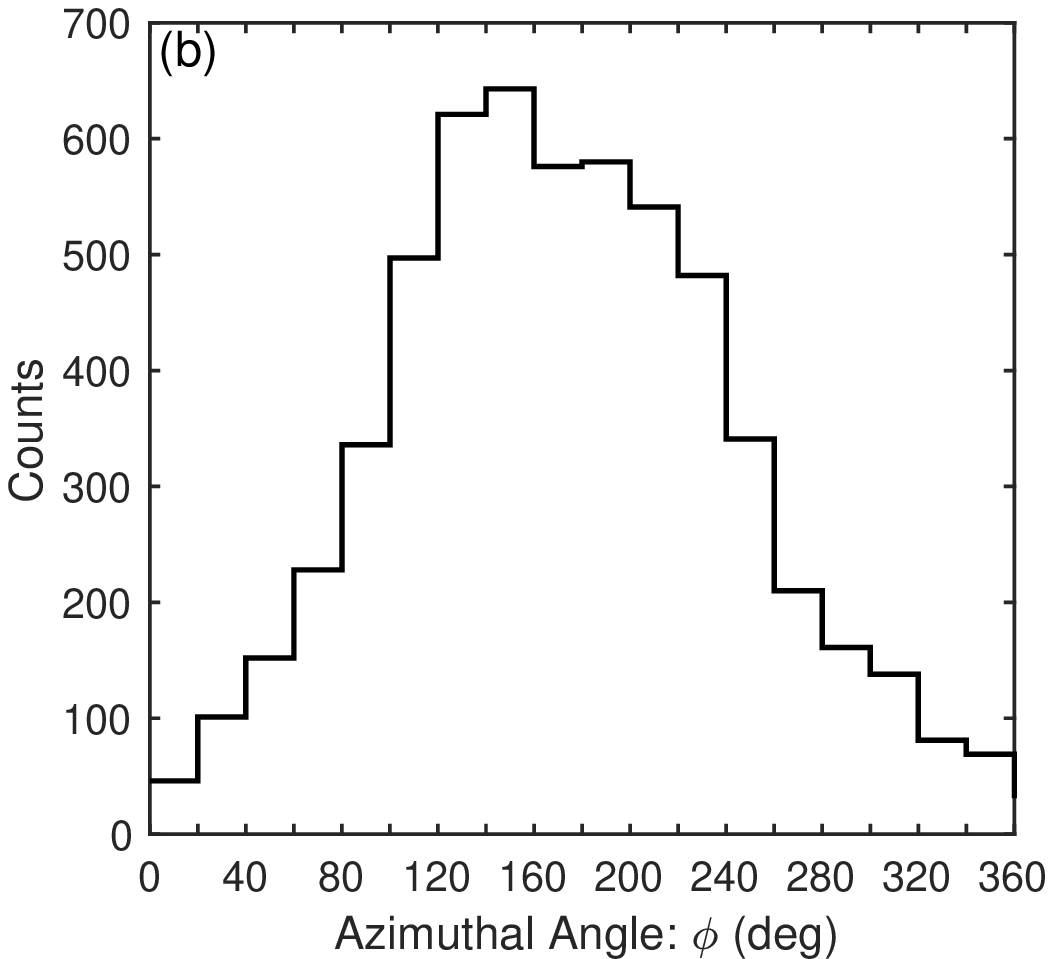}
\caption{Distributions of $z$-axis orientations: (a) the polar angle $\theta$ and (b) the azimuthal angle $\phi$. 
}\label{fig:angle}
\end{figure}

Figure \ref{fig:angle} shows distributions of the orientation of the flux rope central axis, i.e., $z$-axis. The angles $\theta$ and $\phi$ are the polar and azimuthal angles in the RTN coordinates, where R represents the radial direction from the Sun to the PSP, T is the cross product of the solar rotation axis and the R axis, and N follows the right-handed orthogonal rule. These two angles describe the angles of the $z$-axis with respect to N, and its projection onto the RT plane with respect to $R$, respectively. The polar angle of $z$-axis covers almost all angles from 0$^\circ$ to 90$^\circ$ and has gradually more events lying close to the RT plane. The projection of flux rope $z$-axis onto the RT plane has a broad distribution of angles with respect to the R-direction, peaking approximately between $\sim$ 120$^\circ$ and 220$^\circ$. Such a preferred orientation was also found in \cite{Dudok2020} (Figure 2 therein), in which the peak distribution of the azimuthal angle was found to center around  170$^\circ$.

\begin{figure}
\centering
\includegraphics[width=0.45\textwidth]{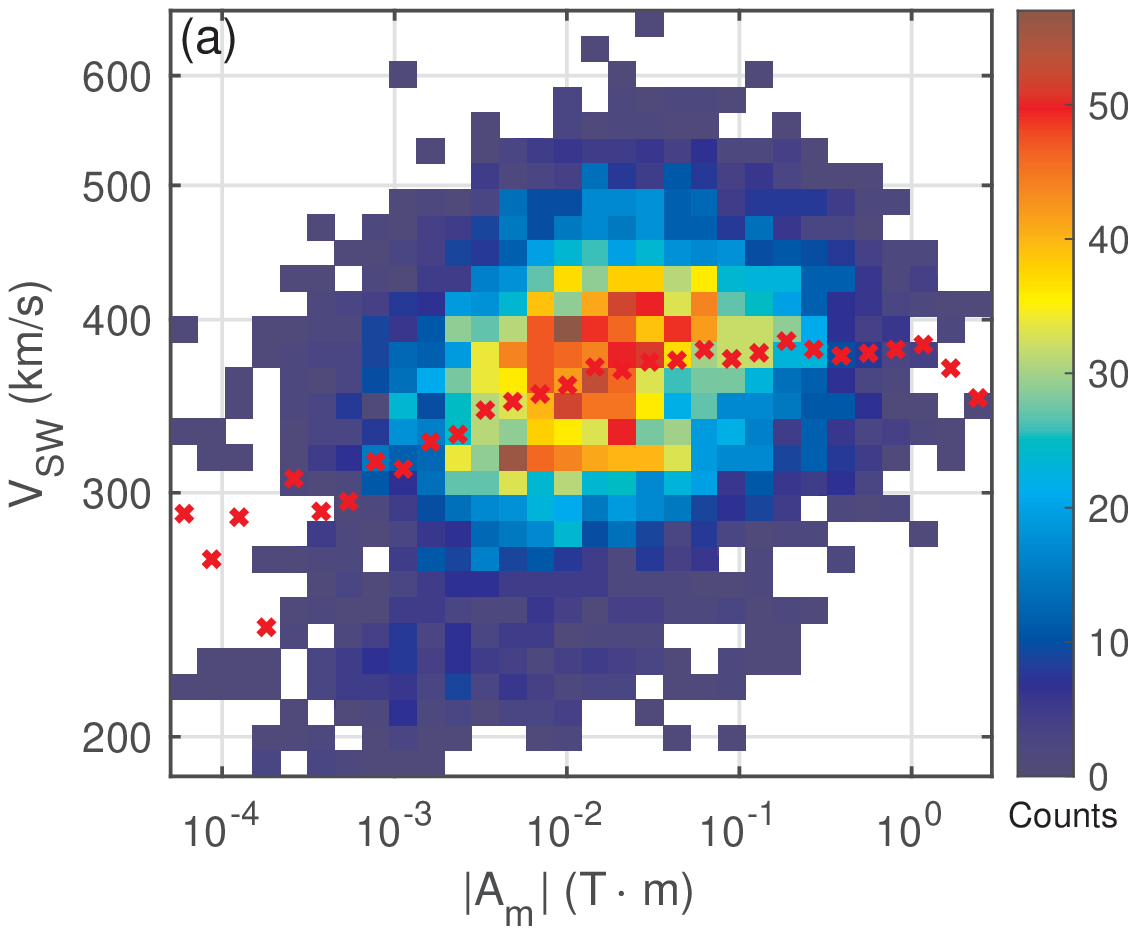}
\includegraphics[width=0.45\textwidth]{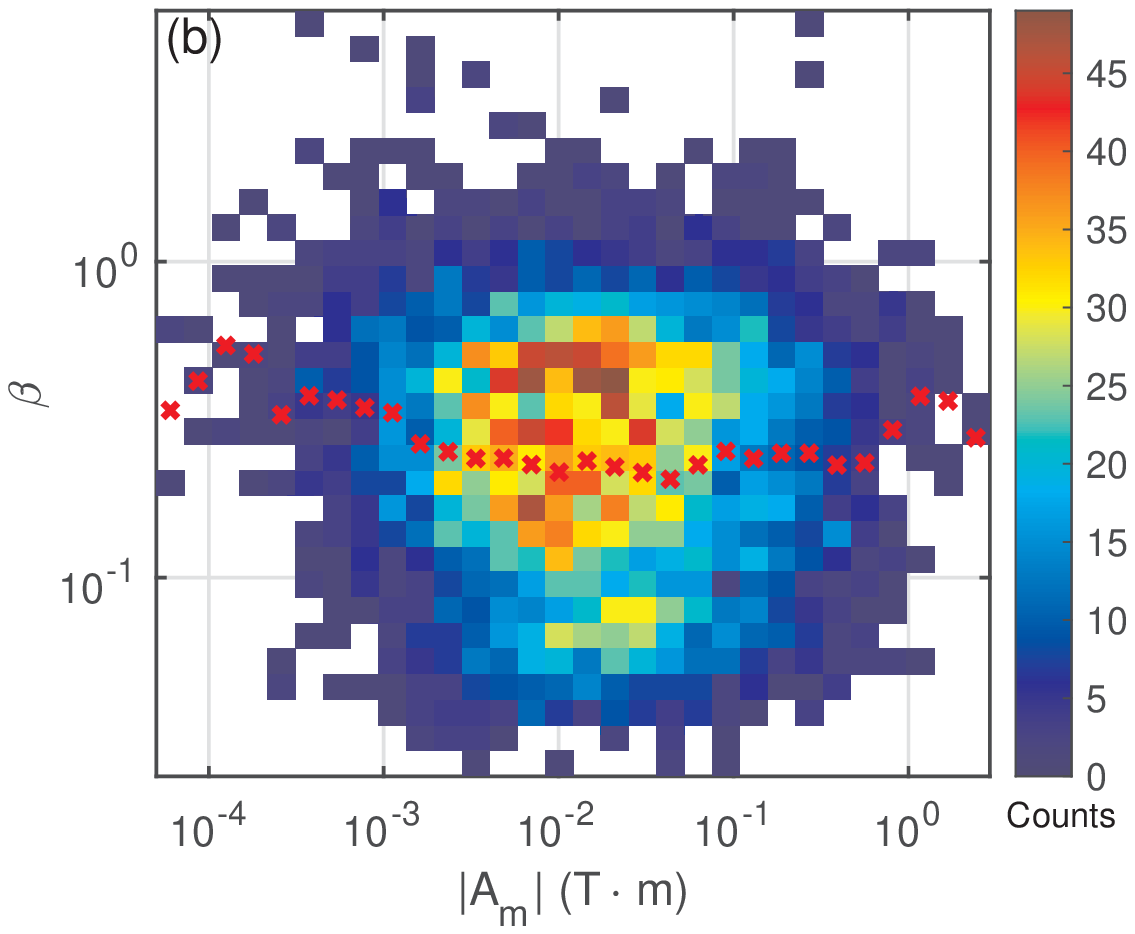}\\
\includegraphics[width=0.45\textwidth]{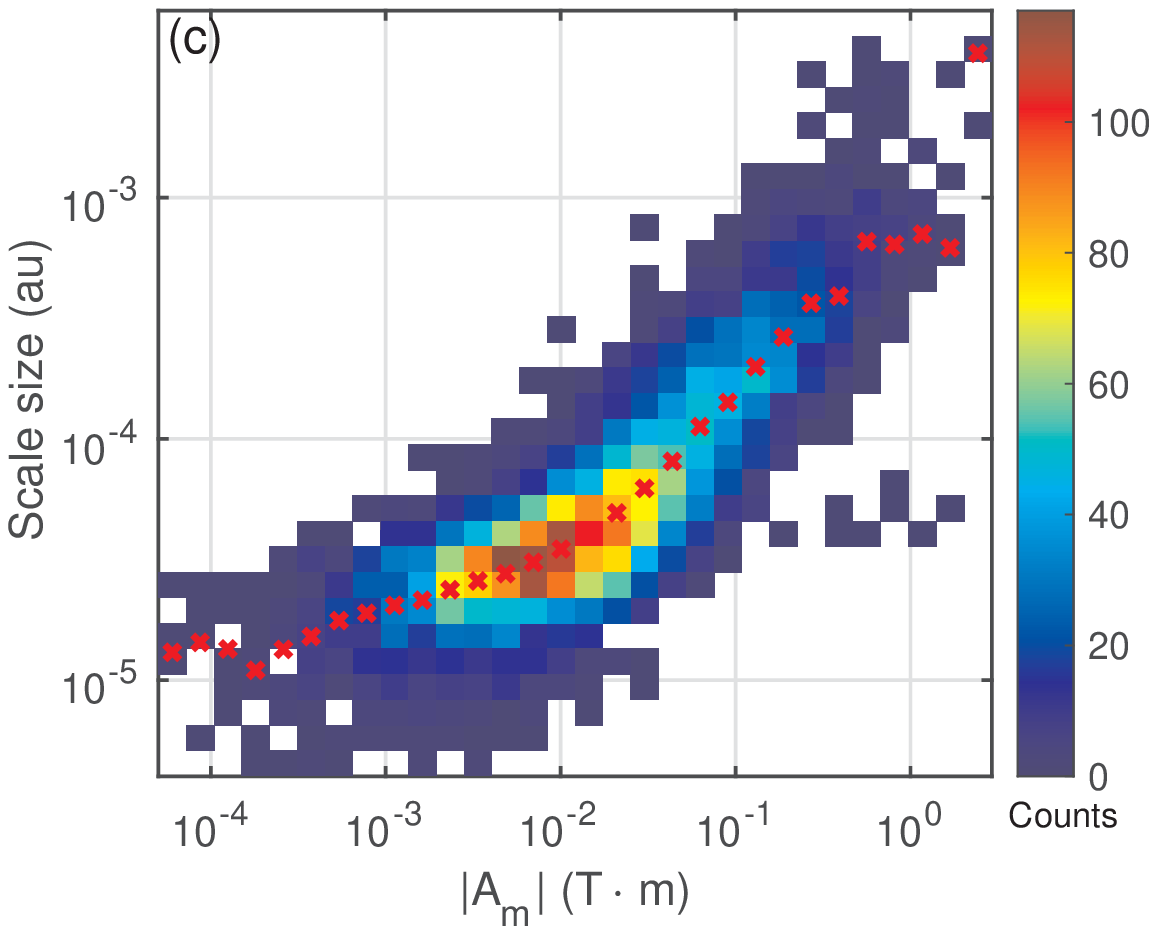}
\includegraphics[width=0.45\textwidth]{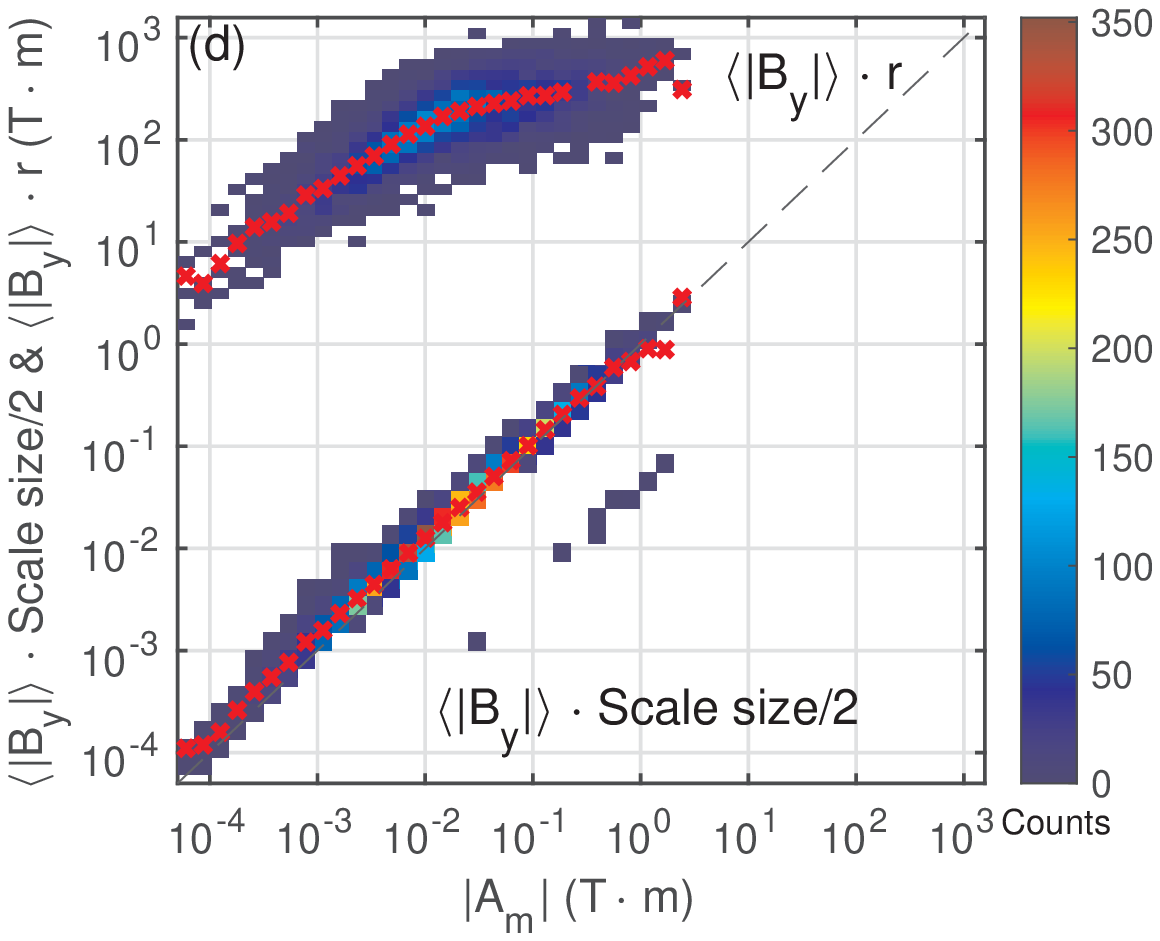}
\caption{2D Distributions of (a) the average solar wind speed $\langle V_{SW}\rangle$, (b) proton $\beta$, (c) scale size, and (d) the products of $\langle |B_y|\rangle$ and radial distance $r$ as well as one half of scale size, together with the corresponding poloidal magnetic flux $|A_m|$. The red crosses represent the average values in each bin of $|A_m|$.
}\label{fig:Am_2D}
\end{figure}

We also examine the correlation between flux rope parameters and $|A_m|$. Figure \ref{fig:Am_2D} presents 2D distributions of various parameters, such as the averaged solar wind speed $\langle V_{SW}\rangle$, proton $\beta$, scale size, the products of $\langle|B_y|\rangle$ and the radial distance $r$ as well as one half of scale size, together with the corresponding poloidal magnetic flux $|A_m|$. Bins with most events cluster near $|A_m|\approx$ 10$^{-2}$ T $\cdot$ m, but each distribution has different tendency for each pair of individual parameters. Figure \ref{fig:Am_2D}(a) presents the solar wind speed averaged within each event interval versus $|A_m|$. Events occur in the solar wind ranging from the rather slow speed, i.e., $\sim$ 187 km s$^{-1}$, to the fast one around 668 km s$^{-1}$. Most events cluster between 300 and 400 km s$^{-1}$ for these first six encounters. 
Figure \ref{fig:Am_2D}(b) displays the relation between the $|A_m|$ and the proton $\beta$, where  $\beta = nk_BT/(B^2/2\mu_0)$, involving the proton density $n$ and temperature $T$ only. Most events tend to have $\beta$ values $\lesssim$ 1. Only a few events have $\beta$ values larger than 1. Notice that this is the distribution for the proton $\beta$ only. One may estimate that the plasma $\beta$ values will increase when including additional contributions to plasma pressure from electrons and alpha particles.
Such tendencies indicate that the magnetic pressure inside most events may dominate over the thermal pressure. The overall tendency is that the averaged solar wind speed and the proton $\beta$ have slight variations with increasing $|A_m|$. Figure \ref{fig:Am_2D}(c) presents the distribution of scale size versus $|A_m|$. Generally, the scale size of most events locates from 10$^{-5}$ to 10$^{-3}$ au, while $|A_m|$ distributes mainly from 10$^{-3}$ to 1 T $\cdot$ m. Such ranges demonstrate that these events are rather small in terms of their spatial scale sizes and amount of flux. The overall trend presents a positive correlation, i.e., larger events tend to have larger poloidal magnetic flux. 
Figure \ref{fig:Am_2D}(d) shows the two products involving the average magnetic field component $\langle |B_y|\rangle$, one as a proxy to poloidal magnetic flux per unit length, versus $|A_m|$, which are well separated in this plot. The top fraction is the product of $\langle |B_y|\rangle$ and the radial distance $r$ where an event is detected.  The other fraction is obtained by multiplying $\langle |B_y|\rangle$ and one-half of scale size. It correlates well with $|A_m|$ because they are intimately related through equation (\ref{eq:eq2}). In other words, they are expected to fall along the diagonal line as shown. In contrast, the average values in the top fraction do not seem to follow a line parallel to the diagonal line, which implies that a radial change of scale size proportional with $r$ is not likely.

\begin{table}[ht]
\begin{center}
\caption{List of SFRs and associated properties during the first six PSP encounters.}
\begin{tabular}{lcc}
\toprule
Column & Label & Explanation\\
\midrule
1 & No. & Event index number\\
2 & Start Time & Event start time; mm/dd/yyyy hh:mm:ss\\
3 & End Time & Event end time; mm/dd/yyyy hh:mm:ss\\
4 & Duration & Event duration (second)\\
5 & $\langle B\rangle$ & Magnetic field strength averaged in event interval (nT)\\
6 & $\langle \beta_p\rangle$ & Average proton beta\\
7 & $\langle V_{SW}\rangle$ & Average solar wind speed (km/s)\\
8 & $\theta$ & Polar angle (deg)\\
9 & $\phi$ & Azimuthal angle (deg)\\
10-12 & z-axis$_{0,1,2}$ & Flux rope $z$-axis components in RTN coordinates\\
13 & size & Flux rope scale size (10$^{-5}$ au)\\
14 & Walen Test Slope & Wal\'en test slope\\
15 & $\langle M_A\rangle$ & Average Alfv\'enic Mach number\\
16 & RD & Heliocentric distance at which the event is identified (au)\\
17 & Cross Helicity & Density of normalized cross helicity $\sigma_c$\\
18 & Residue Energy & Density of normalized residual energy $\sigma_r$\\
19 & $A_m$ & The extreme value of the magnetic flux function (T $\cdot$ m)\\
\bottomrule
\end{tabular}
\tablenotetext{}{(This table is available in its entirety in machine-readable form.)}
\label{table:list}
\end{center}
\end{table}

Table \ref{table:list} presents a brief description of the event list attached to this paper. It includes the start and end times of each event interval in UT, the event duration, the average magnetic field strength, the average proton beta, the average solar wind speed, the polar angle and the azimuthal angle of the flux rope $z$-axis and its three components in RTN coordinates, the flux rope scale size, the Wal\'en test slope, the average Alfv\'en Mach number, the heliocentric distance at which an event is identified, the densities of normalized cross helicity and residual energy, and the extreme value of the magnetic flux function $A_m$.

\section{Case Studies: Configurations of SFR}\label{sec:case}

The interchange reconnection happens between the closed and open magnetic field lines. Such a process may produce magnetic switchbacks \citep{Kasper2019,Bale2019}. Furthermore, this reconnection process was shown by \citet{Drake2020} to be able to generate magnetic flux ropes, which exhibit signatures of magnetic field reversals when crossed by a spacecraft. They identified the observational signatures of possible SFRs within a magnetic switchback interval on 2018 November 5, from 05:45:54 to 05:47:38 UT, which lasted for less than 2 minutes. Notice that our previous study \citep{chen2021} has a lower limit of duration of 5.6 minutes. We can only deduce that the co-existence between switchback and SFR intervals may also be applicable to smaller structures, i.e., with duration down to a few seconds. Now the new detection reported in this study enables us to have a direct comparison for shorter duration events.

\begin{figure}
\centering
\includegraphics[width=0.9\textwidth]{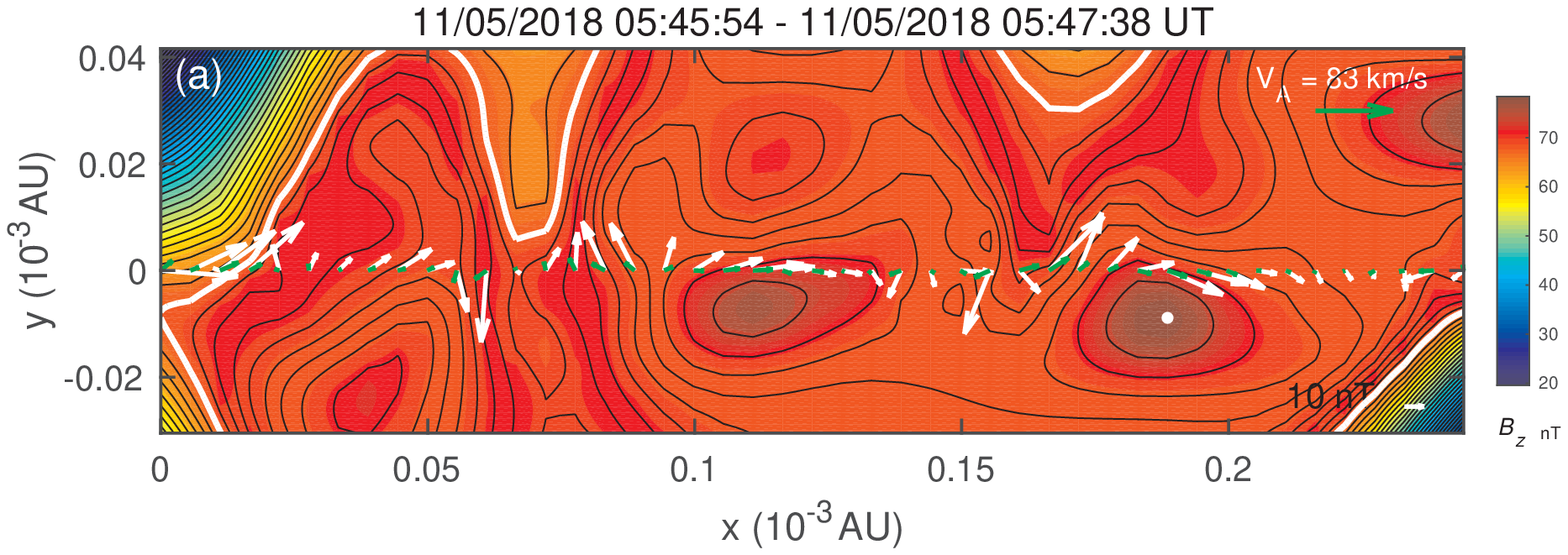}
\includegraphics[width=0.4\textwidth]{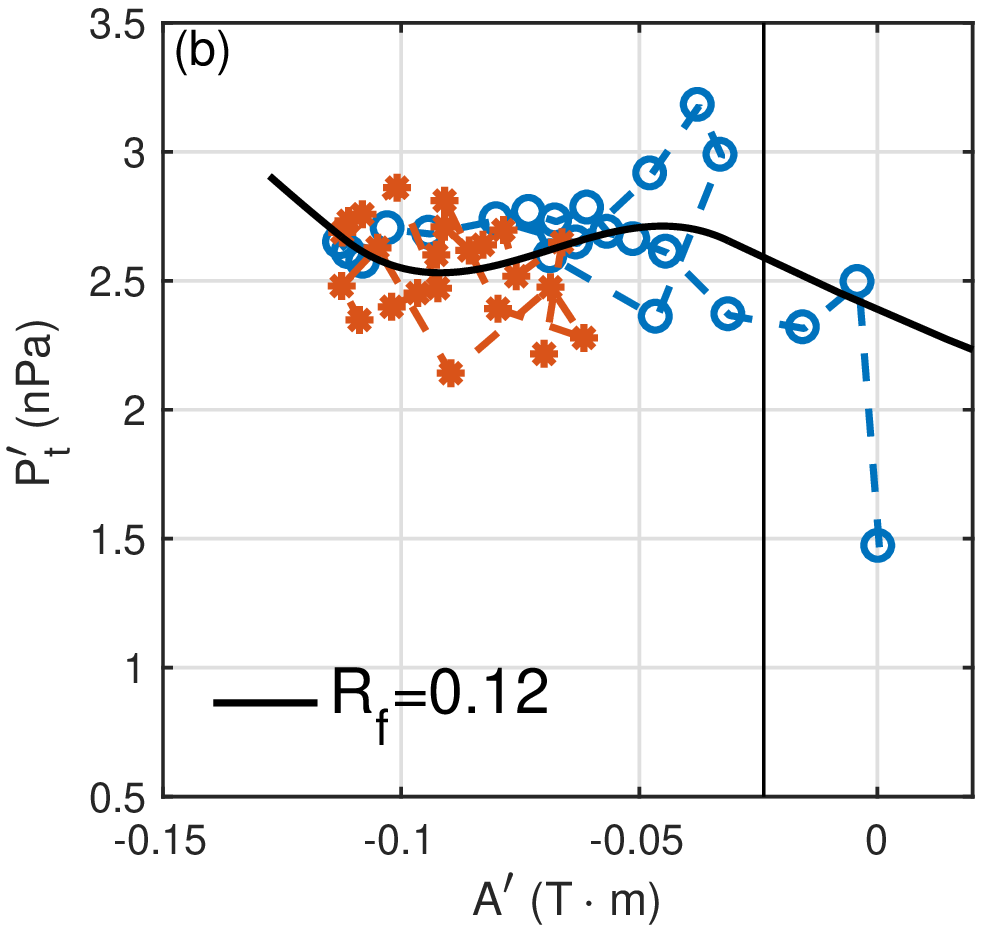}
\includegraphics[width=0.4\textwidth]{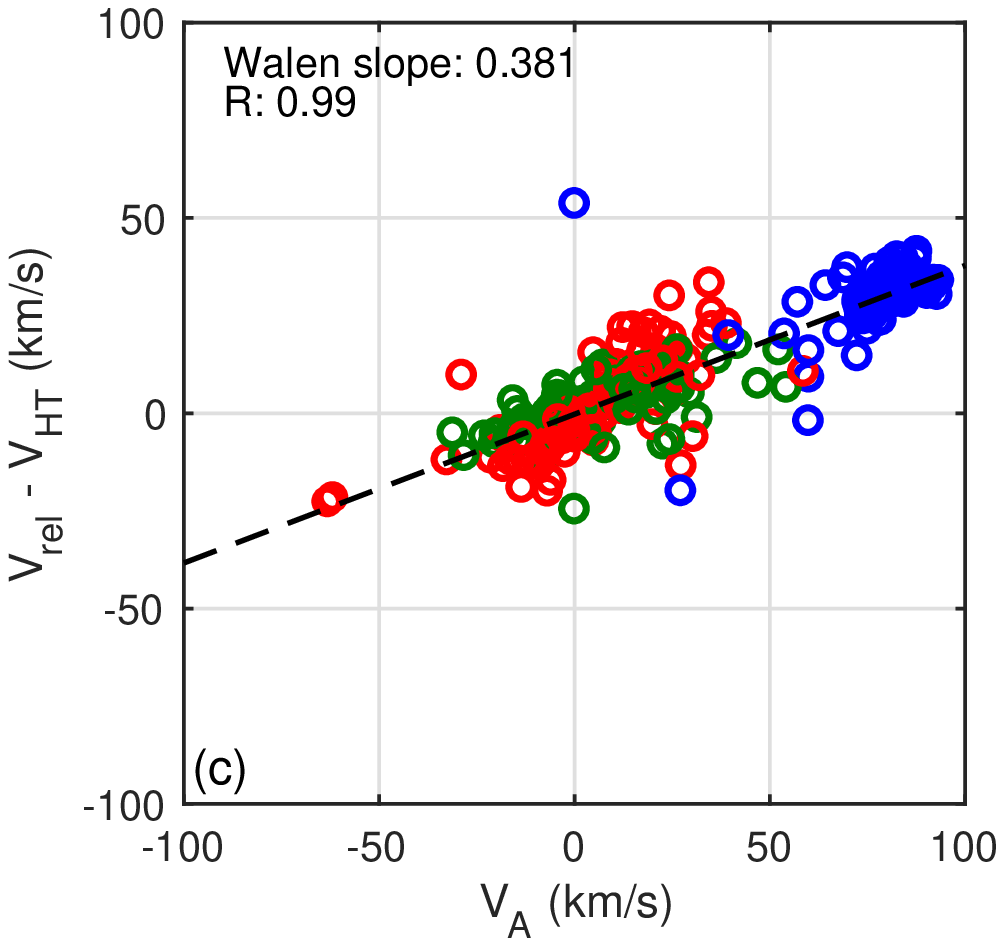}
\caption{Reconstruction results via the new type of GS reconstruction. The panel (a) presents 2D cross-section map from 2018 Nov 5, 05:45:54 - 05:47:38 UT with $\hat{\mathbf{z}}=[0.231, 0.121, 0.965]$ in the RTN coordinates. 
The color background and black curves denote the axial magnetic field $B_z$ and the transverse field $\textbf{B}_t$. The white and green arrows denote the $\textbf{B}_t$ and the remaining flow vectors along the spacecraft path, respectively. Panel (b) displays the $P'_t~versus~A'$ curves from which the cross-section map is reconstructed. The blue circles and red stars denote in-situ measurements on the inbound and outbound paths. The black curve denotes the fitting curve with the fitting residue $R_f$ indicated. The vertical line marks the flux function value corresponding to the white contour in (a). Panel (c) presents the Wal\'en relation between the two velocities. Three components in RTN are denoted by circles in red, green, and blue, respectively. The dashed line represents a linear regression with the slope denoted, together with $R$ representing the correlation coefficient between the two velocities.}\label{fig:3case}
\end{figure}

We first reconstruct the whole switchback interval denoted above by assuming that possible multiple flux ropes have one similar $z$-axis. Figure \ref{fig:3case}(a) shows the 2D cross-section map of the magnetic field configuration. Multiple closed transverse magnetic field line regions and the gradient of the unipolar axial field $B_z$ confirm the existence of multiple flux ropes. The transverse magnetic field and the remaining flow vectors along the spacecraft path are denoted by white and green arrows, respectively. At the beginning and the end of this interval, two sets of vectors are completely reversed and the axial field $B_z$ remains positive, which complies with the signatures of switchback (or spike) boundaries. Moreover, these two sets of vectors seem to be aligned with each other along the entire spacecraft path at $y=0$. The magnitude of the remaining flow, however, tends to be different within each flux rope interval. For example, the green vectors across the first flux rope are rather small when compared with the average Alfv\'en speed for the whole interval, 83 km s$^{-1}$. Such a magnitude indicates a small Wal\'en test slope, i.e., corresponding to a quasi-static flux rope. In the second or third flux rope interval, these vectors become large as indications of Alfv\'enic structures. Figure \ref{fig:3case}(b) presents the two sets of data points corresponding to the two branches and the fitted $P'_t(A')$ curves from which the cross-section map is reconstructed. Figure \ref{fig:3case}(c) presents the Wal\'en relation between the two velocities. Although the Wal\'en test slope may become large in some segments, for the whole interval the Alfv\'enicity remains modest, with a Wal\'en slope 0.381. Moreover, the correlation coefficient is 0.99, which indicates a good alignment between the remaining plasma flow and the local magnetic field.

\begin{figure}
\centering
\includegraphics[width=0.33\textwidth]{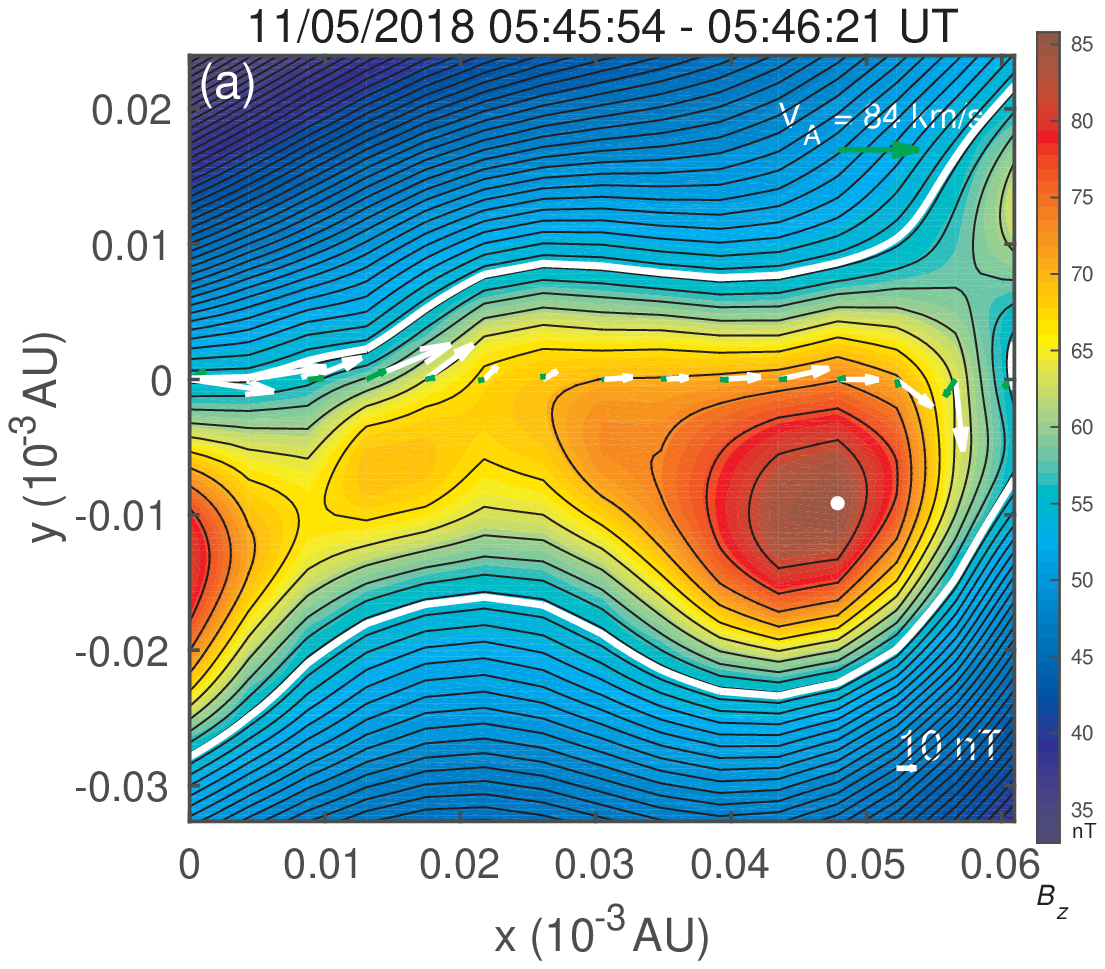}
\includegraphics[width=0.3\textwidth]{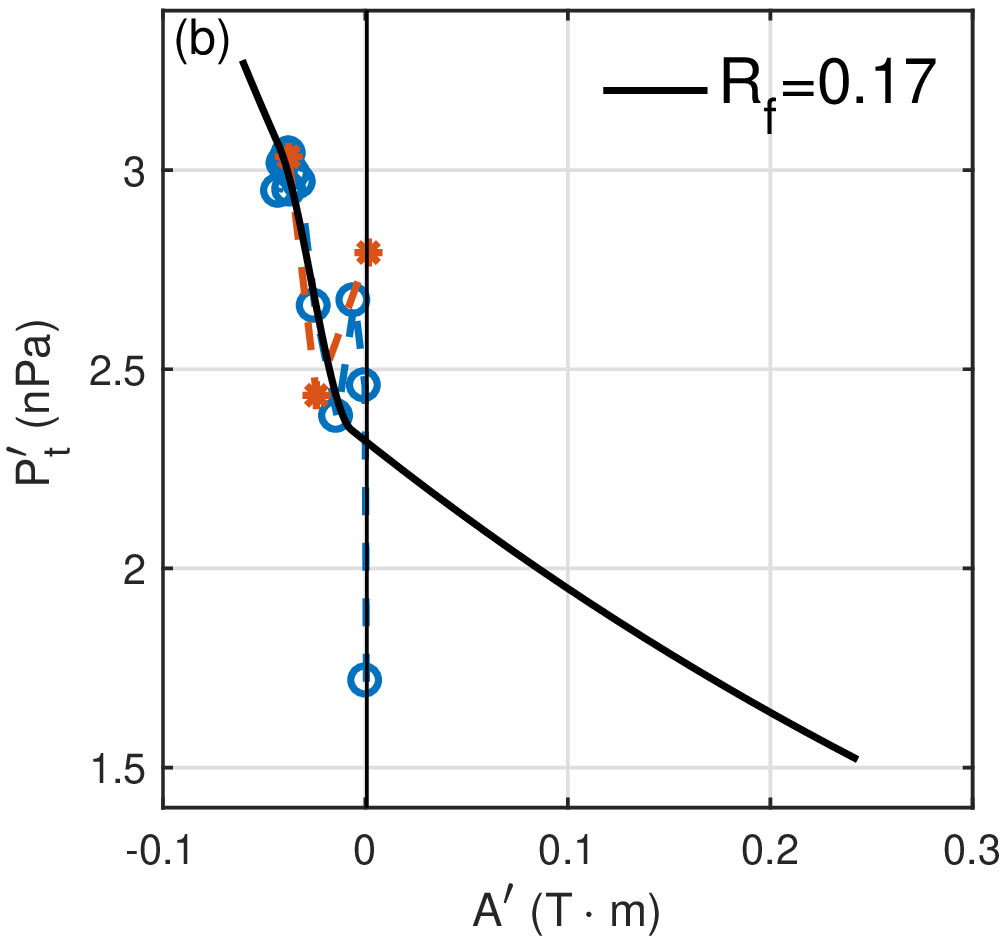}
\includegraphics[width=0.3\textwidth]{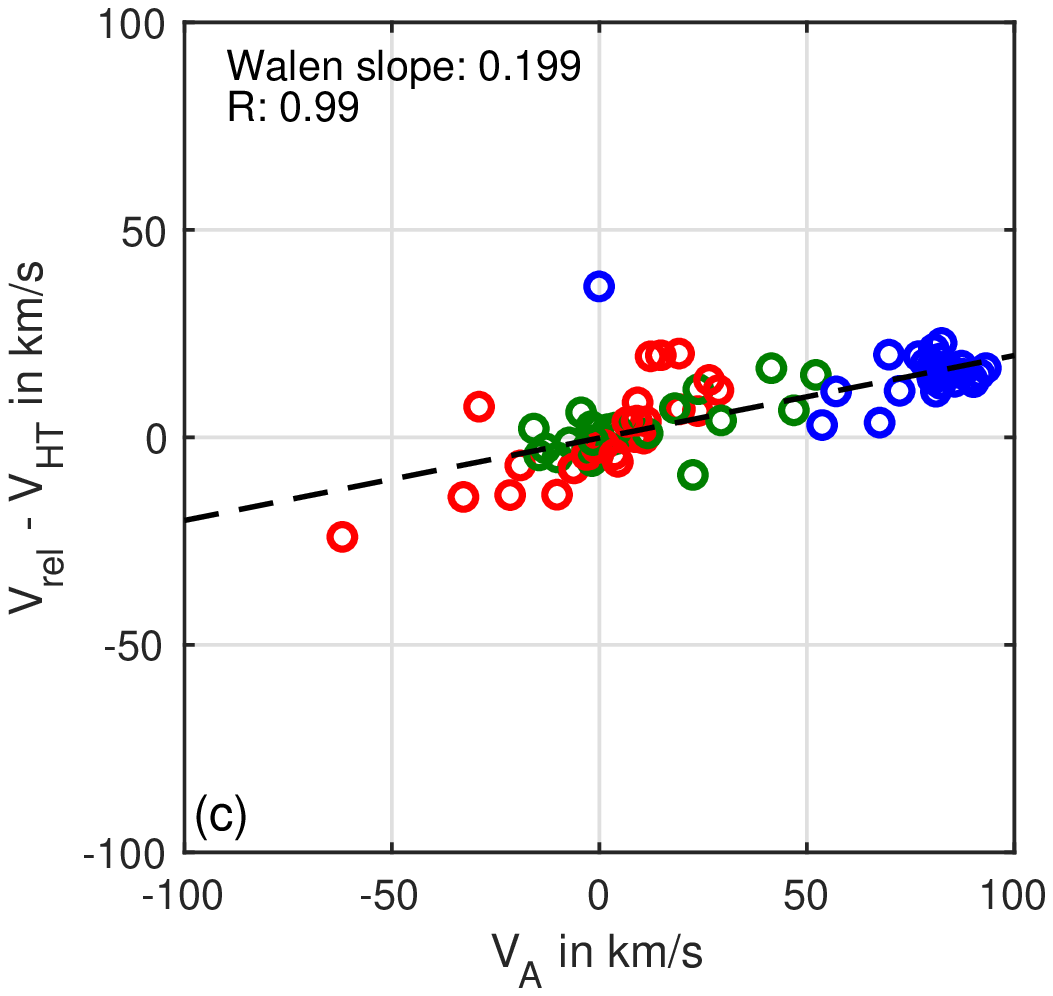}
\includegraphics[width=0.32\textwidth]{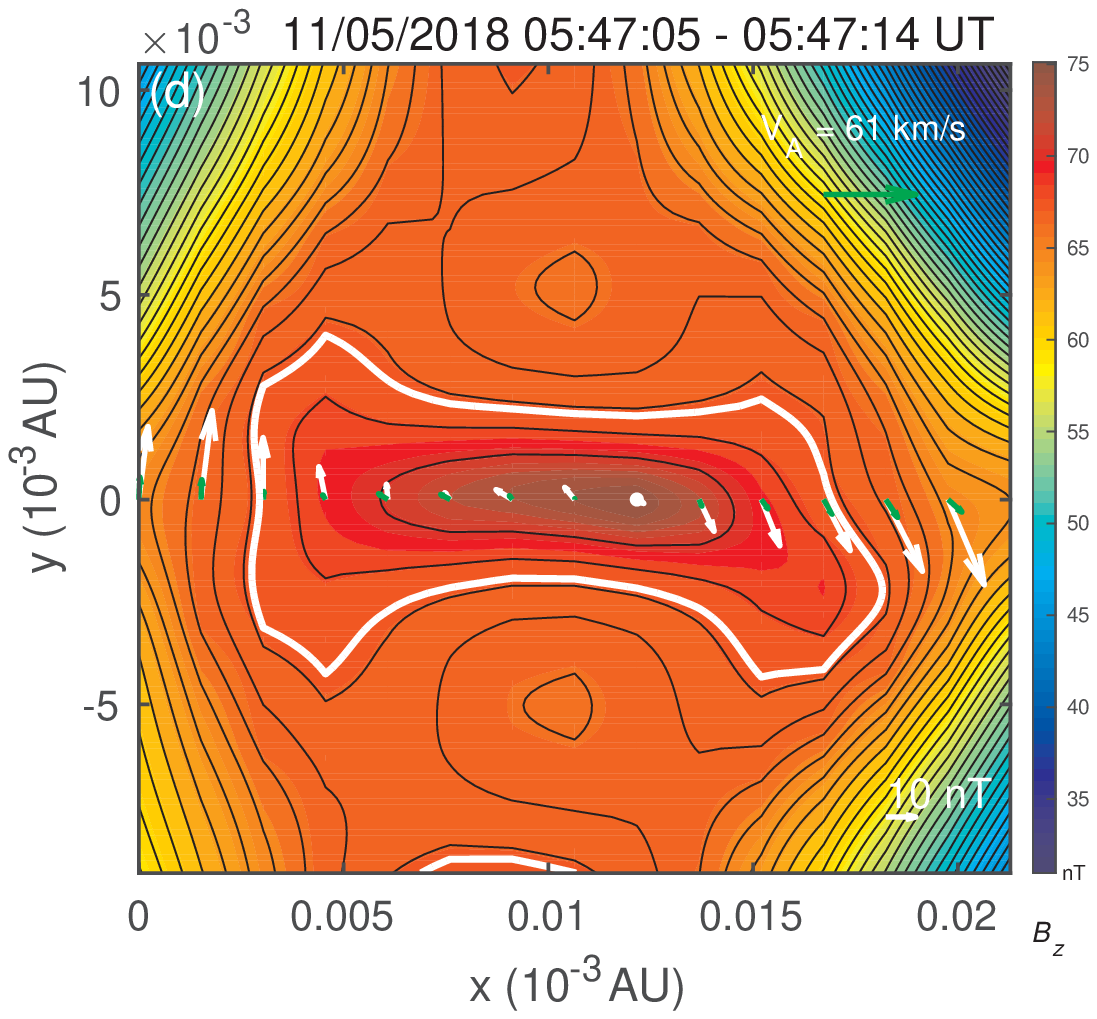}
\includegraphics[width=0.3\textwidth]{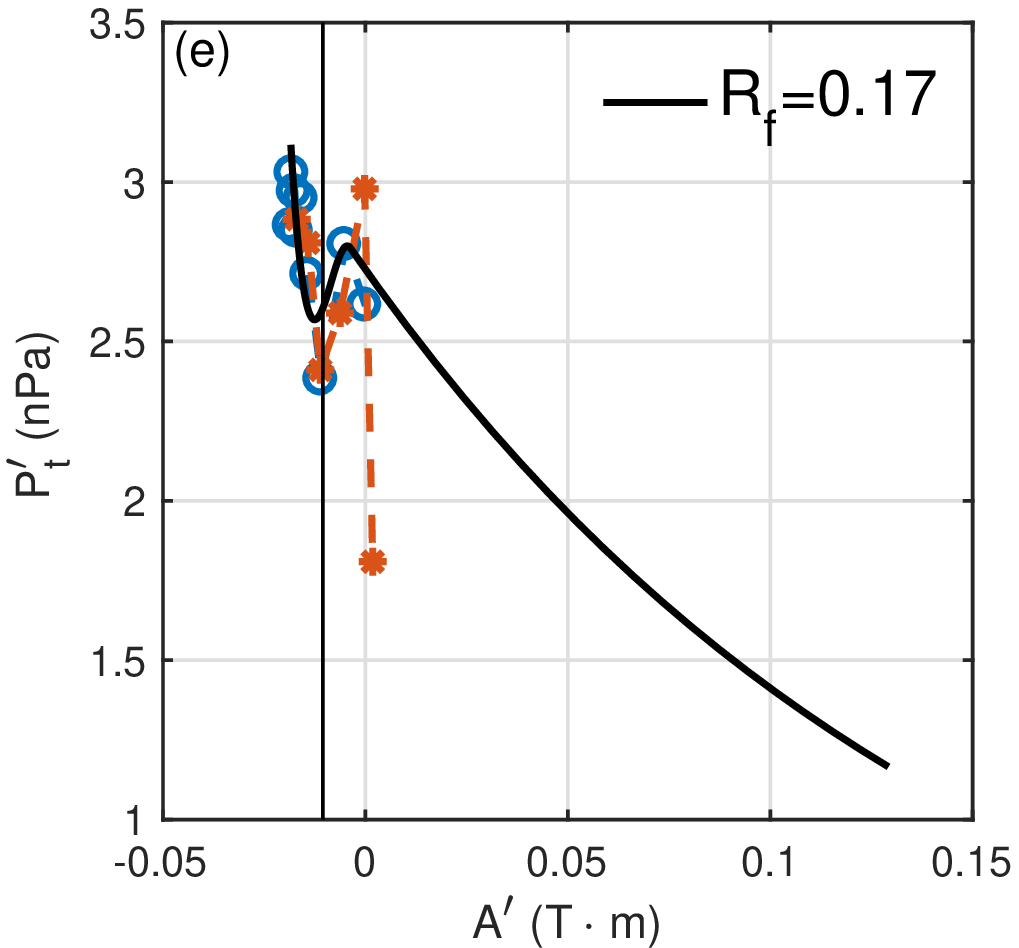}
\includegraphics[width=0.3\textwidth]{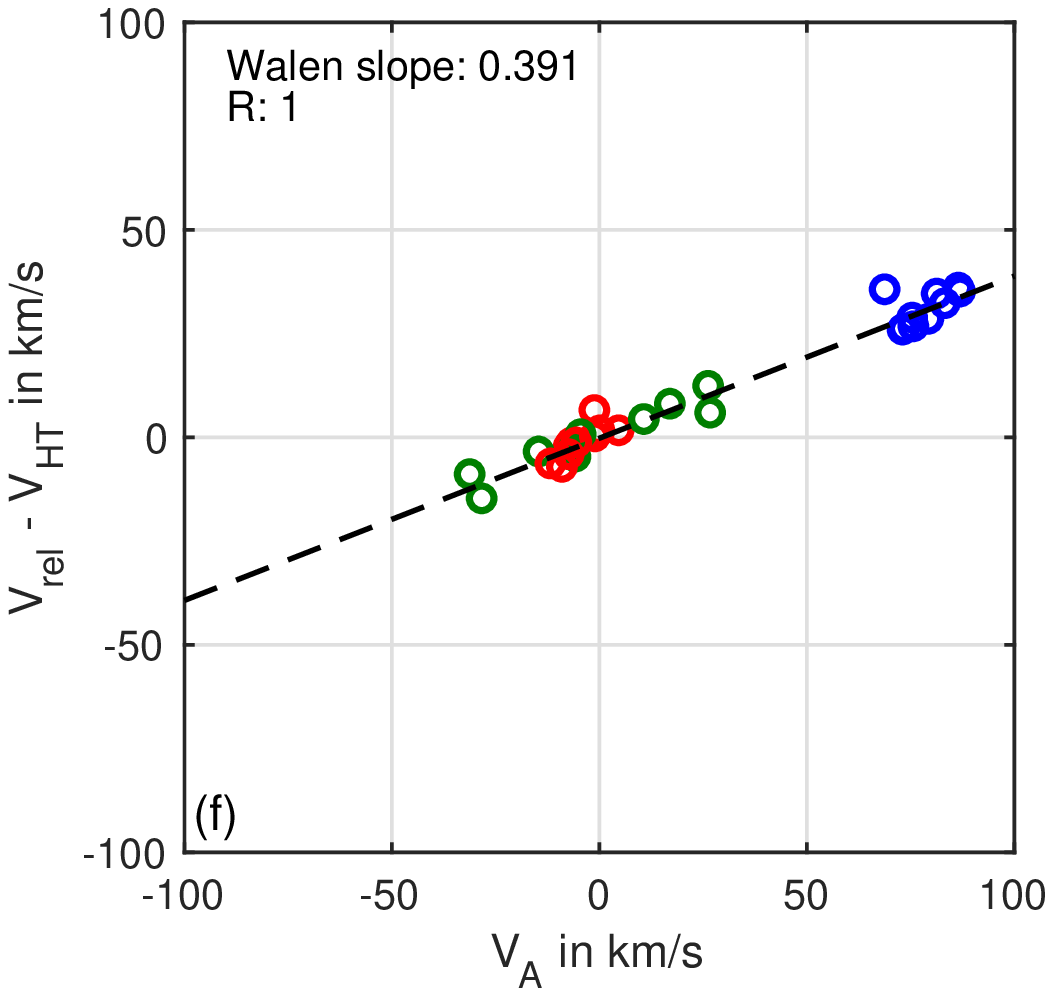}
\caption{Reconstruction results for two sub-intervals (separate SFR events) enclosed by the whole interval given in Figure \ref{fig:3case}. Panels (a-c): event from 05:45:54 to 05:46:21 UT with $\hat{\mathbf{z}}=[0.32, 0.031, 0.947]$. Panels (d-f): event from 05:47:05 to 05:47:14 UT with $\hat{\mathbf{z}}=[-0.009, -0.017, 0.998]$. For each event, the format follows that of Figure \ref{fig:3case}.
}\label{fig:case2}
\end{figure}

\begin{figure}
\centering
\includegraphics[width=0.5\textwidth]{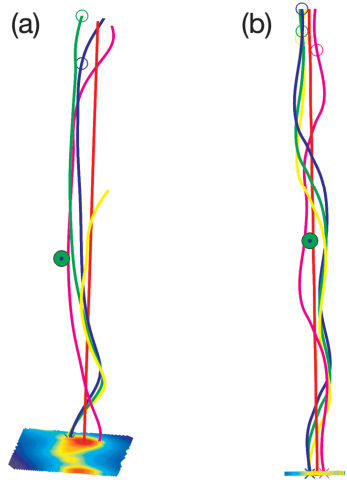}
\caption{3D views of the field line configurations for the events presented in Figure \ref{fig:case2}. The view is along the spacecraft path toward the Sun (the $N$ direction is largely pointing upward). The colorful curves represent the magnetic field lines with the cross-section map displayed on the bottom plane. The round solid green dot marks the spacecraft path.}\label{fig:3D}
\end{figure}

As aforementioned, in order to present the overall structure for the whole switchback interval, Figure \ref{fig:3case} is obtained by assuming that those flux ropes have the same $z$-axis. Now, we perform the new GS-type reconstruction for two individual sub-intervals. Figure \ref{fig:case2} shows results for two intervals close to the first and third flux rope intervals identified by \citet{Drake2020}. Cross-section maps in panels (a, d) and double-folding patterns in panels (b, e) demonstrate flux rope configurations. In panels (c, f), the Wal\'en relation again indicates that these two flux ropes have different levels of Alfv\'enicity, as indicated by the magnitudes of the corresponding Wal\'en slopes.

Figure \ref{fig:3D} displays the corresponding field line configurations in a 3D view toward the Sun. 
Such a configuration is derived from the GS reconstruction results \citep{Hu2001}. Since all three field components are known (in the co-moving frame and due to invariance in the $z$ dimension), one can generate the corresponding values $B_x$, $B_y$, and $B_z$ in a 3D cuboid and thus obtain magnetic field lines in a 3D view.
In this view, the spacecraft path is directly pointing into the plane of sky along the dot. The N direction points vertically upward. The magnetic field lines twist along the $z$-axis, lying on distinct cylindrical surfaces, and thus form the projection of closed field line regions as presented in the 2D cross-section plots in the plane perpendicular to the $z$-axis. The flux rope $z$-axes are not the same, but largely along the N direction. The strong axial fields yield the largest components in the N direction for the two events. Notice that the event interval for Figure \ref{fig:3D}(b) owns the modest Alfv\'enicity and high correlation coefficient between the remaining flow and the magnetic field (see Figure \ref{fig:case2}). Therefore, the twisted magnetic field lines in Figure \ref{fig:3D}(b) also represent the streamlines as viewed in the HT frame. The same applies to Figure \ref{fig:3D}(a), although the remaining flow is small in magnitude.

\begin{figure}
\centering
\includegraphics[width=0.33\textwidth]{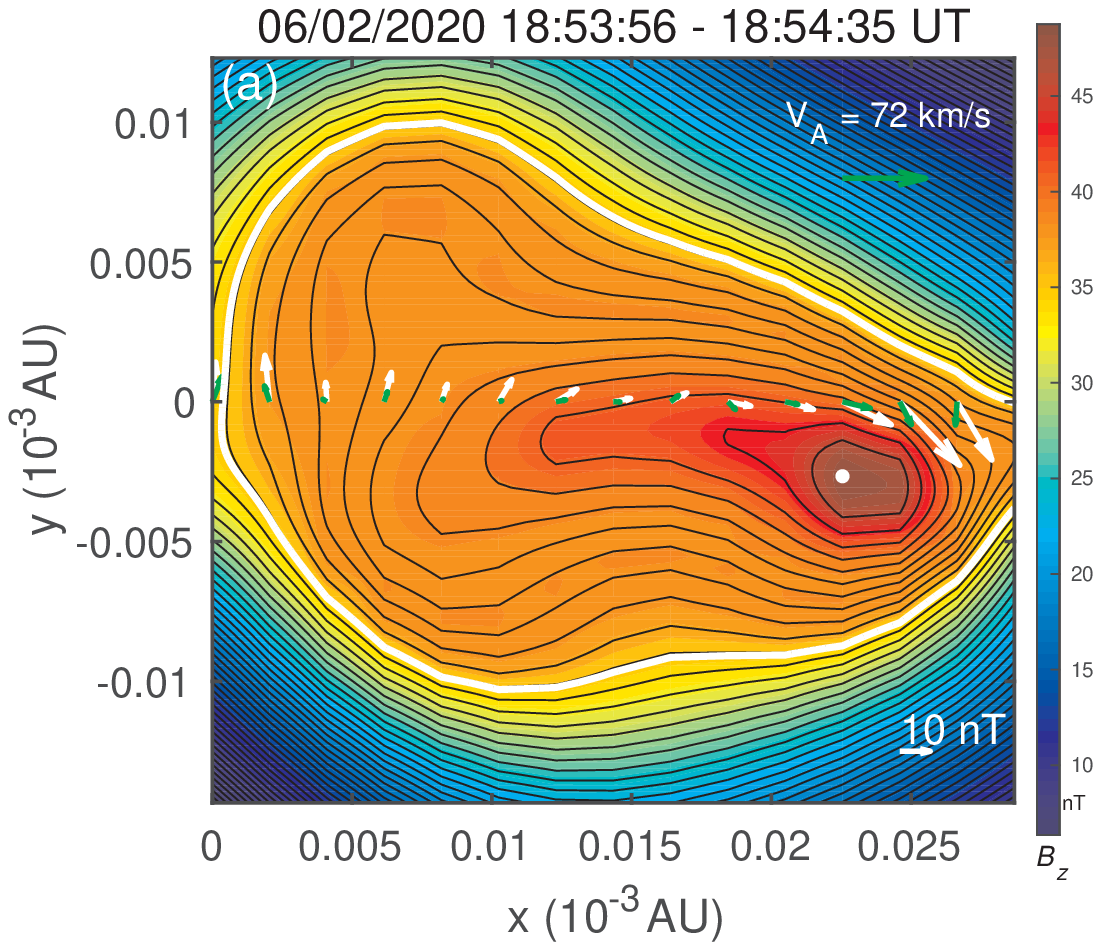}
\includegraphics[width=0.3\textwidth]{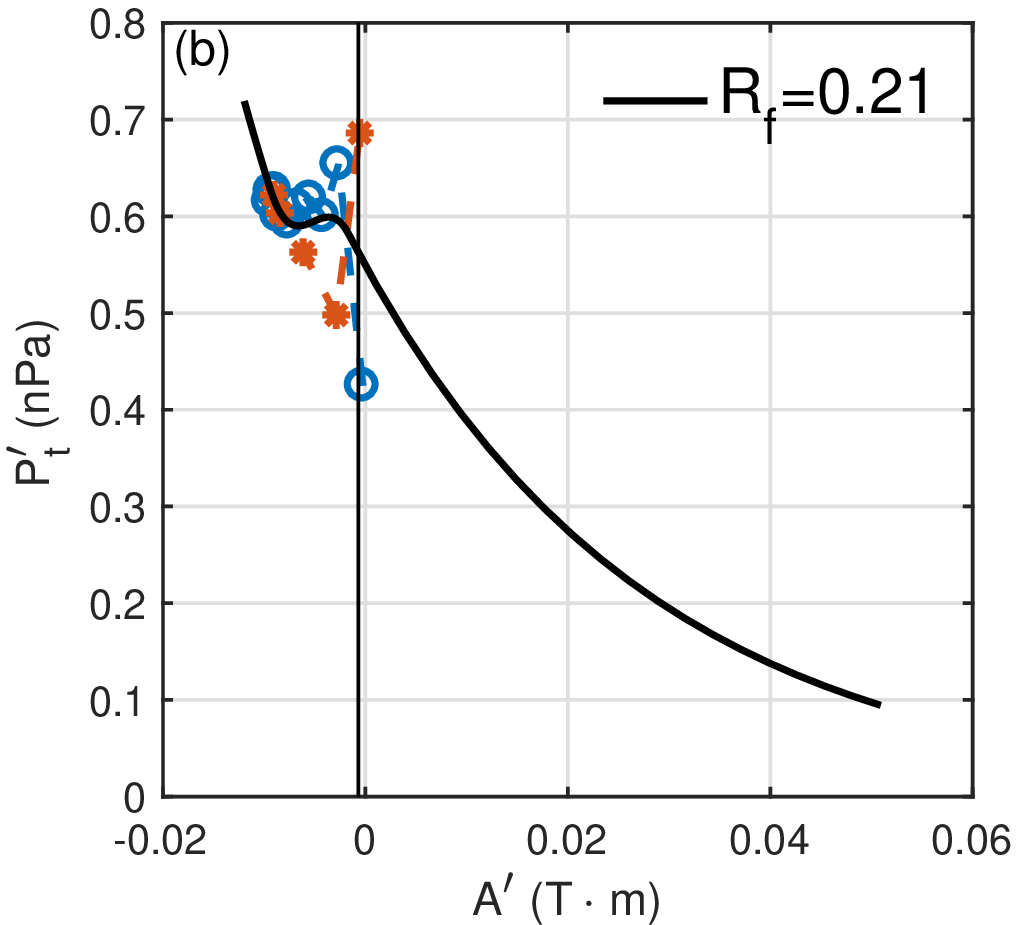}
\includegraphics[width=0.3\textwidth]{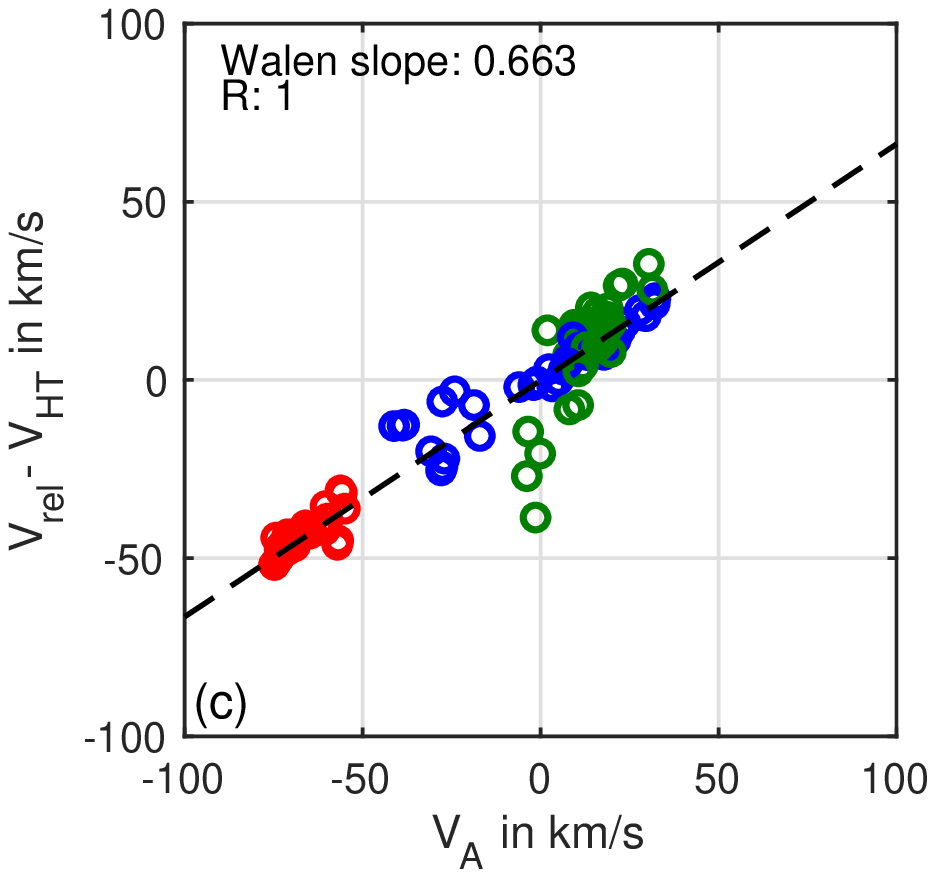}
\caption{New GS-type reconstruction results for an SFR event from 2020 June 2, 18:53:56 to 18:54:35 with $\hat{\mathbf{z}}=[-0.93, 0.34, 0.17]$ in RTN. The format follows that of Figure \ref{fig:3case}.}\label{fig:case3}
\end{figure}

\begin{figure}
\centering
\includegraphics[width=0.4\textwidth]{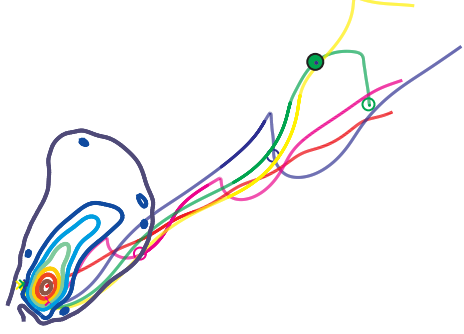}
\caption{3D view of the field line configuration for the event presented in Figure \ref{fig:case3}. The format follows that of Figure \ref{fig:3D}.}\label{fig:3D1}
\end{figure}

In addition to those events reported in \citet{Drake2020}, we also select one additional case from our event list to show the variability in field-line configurations. Figures \ref{fig:case3} and \ref{fig:3D1} exhibit the corresponding reconstruction results, the Wal\'en relation, and the 3D field line view for the event from 2020 June 2, 18:53:56 to 18:54:35 UT. The duration is 40 seconds, and the scale size is 2.93$\times 10^{-5}$ au (about 4383 km). Again, the closed transverse field lines and unipolar axial field verify the configuration of a flux rope, which does not differ from any static SFR structure. Such a flux rope has a rather small scale and contains field-aligned plasma flow with relatively high Alfv\'enicity ($\langle M_A \rangle$ = 0.68 and the Wal\'en slope 0.66). The remaining flow vectors along the spacecraft path are comparable to the average Alfv\'en speed. The 3D view of this event also exhibits an evident knottedness in both magnetic fields (streamlines).

\section{Summary and Discussion}\label{sec:sum}

By applying the extended GS-based algorithm to the PSP in-situ measurements in the first six encounters, we have detected nearly six thousand small-scale magnetic flux ropes including both static structures and those carrying significant field-aligned plasma flows. The duration of these events ranges from 10 to 3,697 seconds, and the scale size is as small as 10$^{-6}$ au. We examine the Alfv\'enicity of these structures with respect to the background radial magnetic field and show distributions of the normalized cross helicity, the normalized residual energy, and the sign of magnetic helicity. The results indicate that most dynamic SFRs have modest to high Alfv\'enicity and propagate anti-sunward. The identified SFRs do not exhibit a preferential sign of magnetic helicity. We also present the macroscopic properties of these structures, such as the distributions of duration, scale size, and $z$-axis orientation angles. The correlations of selected parameters to the poloidal magnetic flux per unit length are displayed via the 2D histograms. Moreover, we carry out GS-type reconstruction to show 2D magnetic field configurations for selected cases. They are composed of spiral field lines which also represent streamlines in the co-moving frame of reference. The results indicate the correspondence between the small-scale flux ropes and magnetic switchbacks. The major findings are summarized as follows.

\begin{enumerate}

\item Most SFR structures possess positive signs of the normalized cross helicity (peak at $\sim$ 1) in the first three encounters, while the background radial field $B_R$ is largely negative. For E4-E6, such a dominance changes with the change of polarity of the magnetic field after HCS crossings. The distributions of normalized residual energy density have different preferences that vary from -1 to 0.

\item The Wal\'en test slopes and cross helicity of most events have the same signs, although the relationship between the two quantities is not linear. The results indicate that the remaining plasma flows inside these structures possess either positive or negative correlation with the local magnetic field.  

\item The magnetic helicity of identified flux rope structures does not have a preferential sign, which implies that the magnetic field lines twist equally in either a right-handed or left-handed manner, corresponding to the right-handed or left-handed chirality.

\item The flux rope $z$-axis orientation shows an increasing tendency of small inclination angle with respect to the RT plane. There is a broad peak centered around the -R direction between $\sim$ 120$\deg$ and 220$\deg$ in the distribution of the azimuthal angle.

\item The distributions of the poloidal magnetic flux per unit length $|A_m|$, duration, and scale size generally follow power-law functions. The scale size and a proxy of poloidal magnetic flux distributions seem to scale with $|A_m|$.

\item The overlapping of switchback and SFR intervals with duration as small as a few seconds is confirmed via the new GS-type reconstruction. Such overlapping includes both quasi-static and Alfv\'enic SFRs. The latter structure still possesses a configuration of twisted field lines (equivalent to streamlines), characteristic of a magnetic flux rope.

\end{enumerate}

In this paper, we further reveal the variability in the configurations of flux ropes, broadly defined, at close distances from the Sun. They reach temporal scales down to a few seconds. Our results reveal the prevalence of modest to high Alfv\'enicity in these broadly-defined SFR structures. Our extended GS-type reconstruction is able to characterize their configurations under one unified theoretical framework. We find that flux-rope-like structures arise very frequently in the inner heliosphere ($r<0.3$ au), which has a daily occurrence rate of over one hundred in some encounters. Considering the close heliocentric distances at which they were detected and the overlapping with magnetic switchbacks, it is very likely for these flux rope-like structures to be formed via the similar mechanisms, such as the interchange reconnection in low corona, as proposed by \citet{Drake2020}.

Another interesting fact is that while the static SFR was conjectured to be generated via the MHD turbulence at larger heliocentric distances \citep{Zheng2018}, the broadly defined flux rope (i.e., allowing for modest to high level of Alfv\'enicity) in this study may also be tied to turbulence. As demonstrated here, the magnetic field and remaining plasma flow constitute the Alfv\'enic alignment inside this twisted structure. Such an alignment, as a result of rapid relaxation processes intrinsic to ideal MHD, happens locally and is associated with the kinetic energy and pressure gradients \citep{Matthaeus2008,Osman2011}. Moreover, it appears to be random, contains wide changes, and complies with the MHD turbulence description involving the ideal MHD invariants, such as the magnetic helicity and cross helicity, parameters commonly derived from in-situ measurements. Notice that such a correlation does not necessarily require a high correlation coefficient between the magnetic field and velocity. In other words, the Wal\'en test slopes can range from small, modest, to high values, as we present here.

\acknowledgments We acknowledge NASA grants 80NSSC21K0003, 80NSSC19K0276, and 80NSSC18K0622
and NSF grant No. AGS-1650854 for support. We also acknowledge the NASA Parker Solar Probe Mission and FIELDS as well as SWEAP teams led by S. D. Bale and J. Kasper for use of data, which are downloaded from the NASA CDAWeb ({\url{https://cdaweb.gsfc.nasa.gov/index.html/}}). This work was made possible in part by a grant of high performance computing resources and technical support from the Alabama Supercomputer Authority.

\bibliography{bib_database}

\end{document}